\documentclass[
aps,prc,letterpaper,preprintnumbers,superscriptaddress,showpacs,amsmath,letterp
aper,floatfix,groupedaddress,hyperref,nofootinbib,twocolumn]{
revtex4}
\usepackage{multirow}
\usepackage{color}
\usepackage{tabularx}
\usepackage{graphicx}
\usepackage{amsthm}
\usepackage{amsmath}
\usepackage{amssymb}
\usepackage{bm}
\usepackage{textcomp}
\usepackage{url}
\usepackage{tabularx}
\usepackage{verbatim}

\newcommand{\vz}{\mbox{$V_Z$}}
\newcommand{\pt}{\mbox{$p_T$}}

\newcommand{\meanptt}{\mbox{$\langle p_{T}^{2} \rangle $}}

\newcommand{\jpsi}{\mbox{$J/\psi$}}
\newcommand{\bjpsi}{$\boldsymbol{\jpsi}$ }

\newcommand{\psiprim}{\mbox{$\rm{\psi`}$}}
\newcommand{\chic}{\mbox{$\rm{\chi}_{c}$}}

\newcommand{\aaa}{\mbox{A+A}}

\newcommand{\bdau}{\mbox{\bf{{\it d}+Au}}}
\newcommand{\dau}{\mbox{{\it d}+Au}}

\newcommand{\pa}{\mbox{{\it p}+A}}
\newcommand{\pp}{\mbox{{{\it p}+{\it p}}}}
\newcommand{\bpp}{\mbox{\bf{{\it{p}+{\it p}}}}}

\newcommand{\snn}{\mbox{$\sqrt{s_{NN}}$}}
\newcommand{\bsnn}{\mbox{$\boldsymbol{\snn}$} }

\newcommand{\fm}{\mbox{$\mathrm{fm}$}}
\newcommand{\cm}{\mbox{$\mathrm{cm}$}}
\newcommand{\mb}{\mbox{$\mathrm{mb}$}}
\newcommand{\nb}{\mbox{$\mathrm{nb}$}}
\newcommand{\gev}{\mbox{$\mathrm{GeV}$}}
\newcommand{\bgev}{\mbox{$\mathbf{GeV}$}}
\newcommand{\gevc}{\mbox{$\mathrm{GeV/}c$}}
\newcommand{\gevcc}{\mbox{$\mathrm{GeV/}c^2$}}

\newcommand{\nsig}{n\sigma_{e}}
\newcommand{\nsigal}{n\sigma_{\alpha}}

\newcommand{\nsigPi}{n\sigma_{\pi}}
\newcommand{\nsigP}{n\sigma_{p}}

\newcommand{\invbeta}{1/\beta}

\newcommand{\eop}{E/p}

\newcommand{\invbetacut}{|\invbeta-1|<0.03 }

\newcommand{\rda}{\mbox{$R_{\textit{dA}}$}}
\newcommand{\dedx}{\mbox{$dE/dx$}}

\newcommand{\ncoll}{\mbox{$N_{\textrm{coll}}$}}

\newcommand{\mncoll}{\mbox{$\langle N_{\textrm{coll}} \rangle$}}
\newcommand{\mnpart}{\mbox{$\langle N_{\textrm{part}} \rangle$}}

\newcommand{\sigabs}{\sigma_{\textrm{abs}}}
\newcommand{\sigabsresult}{\sigabs = 0.0  {}^{+3.8}_{-0.0} \ \textrm{(stat.)} \ {}^{+2.1}_{-0.0} \ \textrm{(syst.)} \ \mb}
\newcommand{\meanpttresult}{\meanptt = 3.45 \pm 0.85 \ (\textrm{stat.}) \pm 1.22 \ (\textrm{syst.}) \ (\gevc)^{2}}
\newcommand{\meanpttresultdau}{\meanptt = 3.70 \pm 0.33 \ (\textrm{stat.}) \pm 0.44 \ (\textrm{syst.}) \ (\gevc)^{2}}

\graphicspath{ 
  {images/} 
}

\begin{document}
\title{ $\bjpsi$ production at low transverse momentum  in $\bpp$ and $\bdau$ 
collisions \\ at $\bsnn$ = 200 $\bgev$ }
\affiliation{AGH University of Science and Technology, FPACS, Cracow 30-059, Poland}
\affiliation{Argonne National Laboratory, Argonne, Illinois 60439}
\affiliation{Brookhaven National Laboratory, Upton, New York 11973}
\affiliation{University of California, Berkeley, California 94720}
\affiliation{University of California, Davis, California 95616}
\affiliation{University of California, Los Angeles, California 90095}
\affiliation{Central China Normal University, Wuhan, Hubei 430079}
\affiliation{University of Illinois at Chicago, Chicago, Illinois 60607}
\affiliation{Creighton University, Omaha, Nebraska 68178}
\affiliation{Czech Technical University in Prague, FNSPE, Prague, 115 19, Czech Republic}
\affiliation{Nuclear Physics Institute AS CR, 250 68 Prague, Czech Republic}
\affiliation{Frankfurt Institute for Advanced Studies FIAS, Frankfurt 60438, Germany}
\affiliation{Institute of Physics, Bhubaneswar 751005, India}
\affiliation{Indian Institute of Technology, Mumbai 400076, India}
\affiliation{Indiana University, Bloomington, Indiana 47408}
\affiliation{Alikhanov Institute for Theoretical and Experimental Physics, Moscow 117218, Russia}
\affiliation{University of Jammu, Jammu 180001, India}
\affiliation{Joint Institute for Nuclear Research, Dubna, 141 980, Russia}
\affiliation{Kent State University, Kent, Ohio 44242}
\affiliation{University of Kentucky, Lexington, Kentucky, 40506-0055}
\affiliation{Korea Institute of Science and Technology Information, Daejeon 305-701, Korea}
\affiliation{Institute of Modern Physics, Chinese Academy of Sciences, Lanzhou, Gansu 730000}
\affiliation{Lawrence Berkeley National Laboratory, Berkeley, California 94720}
\affiliation{Lehigh University, Bethlehem, PA, 18015}
\affiliation{Max-Planck-Institut fur Physik, Munich 80805, Germany}
\affiliation{Michigan State University, East Lansing, Michigan 48824}
\affiliation{National Research Nuclear Univeristy MEPhI, Moscow 115409, Russia}
\affiliation{National Institute of Science Education and Research, Bhubaneswar 751005, India}
\affiliation{National Cheng Kung University, Tainan 70101 }
\affiliation{Ohio State University, Columbus, Ohio 43210}
\affiliation{Institute of Nuclear Physics PAN, Cracow 31-342, Poland}
\affiliation{Panjab University, Chandigarh 160014, India}
\affiliation{Pennsylvania State University, University Park, Pennsylvania 16802}
\affiliation{Institute of High Energy Physics, Protvino 142281, Russia}
\affiliation{Purdue University, West Lafayette, Indiana 47907}
\affiliation{Pusan National University, Pusan 46241, Korea}
\affiliation{University of Rajasthan, Jaipur 302004, India}
\affiliation{Rice University, Houston, Texas 77251}
\affiliation{University of Science and Technology of China, Hefei, Anhui 230026}
\affiliation{Shandong University, Jinan, Shandong 250100}
\affiliation{Shanghai Institute of Applied Physics, Chinese Academy of Sciences, Shanghai 201800}
\affiliation{State University Of New York, Stony Brook, NY 11794}
\affiliation{Temple University, Philadelphia, Pennsylvania 19122}
\affiliation{Texas A\&M University, College Station, Texas 77843}
\affiliation{University of Texas, Austin, Texas 78712}
\affiliation{University of Houston, Houston, Texas 77204}
\affiliation{Tsinghua University, Beijing 100084}
\affiliation{United States Naval Academy, Annapolis, Maryland, 21402}
\affiliation{Valparaiso University, Valparaiso, Indiana 46383}
\affiliation{Variable Energy Cyclotron Centre, Kolkata 700064, India}
\affiliation{Warsaw University of Technology, Warsaw 00-661, Poland}
\affiliation{Wayne State University, Detroit, Michigan 48201}
\affiliation{World Laboratory for Cosmology and Particle Physics (WLCAPP), Cairo 11571, Egypt}
\affiliation{Yale University, New Haven, Connecticut 06520}

\author{L.~Adamczyk}\affiliation{AGH University of Science and Technology, FPACS, Cracow 30-059, Poland}
\author{J.~K.~Adkins}\affiliation{University of Kentucky, Lexington, Kentucky, 40506-0055}
\author{G.~Agakishiev}\affiliation{Joint Institute for Nuclear Research, Dubna, 141 980, Russia}
\author{M.~M.~Aggarwal}\affiliation{Panjab University, Chandigarh 160014, India}
\author{Z.~Ahammed}\affiliation{Variable Energy Cyclotron Centre, Kolkata 700064, India}
\author{I.~Alekseev}\affiliation{Alikhanov Institute for Theoretical and Experimental Physics, Moscow 117218, Russia}
\author{A.~Aparin}\affiliation{Joint Institute for Nuclear Research, Dubna, 141 980, Russia}
\author{D.~Arkhipkin}\affiliation{Brookhaven National Laboratory, Upton, New York 11973}
\author{E.~C.~Aschenauer}\affiliation{Brookhaven National Laboratory, Upton, New York 11973}
\author{A.~Attri}\affiliation{Panjab University, Chandigarh 160014, India}
% ...

\collaboration{STAR Collaboration}\noaffiliation

\preprint{Submitted to Phys. Rev. C}
\date{\today}
\pagenumbering{arabic}
\pagestyle{myheadings} 
\thanks{}




\begin{abstract}

We report on the measurement of $\jpsi$ production in the dielectron channel at midrapidity ($|y|<1$) in $\pp$ and $\dau$ collisions at $\snn = 200 \ \gev$ from the STAR experiment at the Relativistic Heavy Ion Collider.
The transverse momentum $\pt$ spectra in $\pp$ for $\pt < 4 \ \gevc$ and $\dau$ collisions for $\pt < 3 \ \gevc$ are presented. 
These measurements extend the STAR coverage for $\jpsi$ production in $\pp$ collisions to low $\pt$. 
The $\meanptt$ from the measured $\jpsi$ invariant cross section in $\pp$ and $\dau$ collisions are evaluated and compared to similar measurements at other collision energies.
The nuclear modification factor for $\jpsi$ is extracted as a function of $\pt$ and collision centrality in $\dau$ and  
compared to model calculations using the modified nuclear parton distribution function and a final-state $\jpsi$ nuclear absorption cross section.  

\end{abstract}



\maketitle


\section{Introduction\label{sec:introduction}}

The study of $\jpsi$ production has been extensively used to probe the medium 
created in relativistic heavy-ion collisions, where a transition 
from hadronic matter to a deconfined quark gluon plasma (QGP) takes place
~\cite{ref:rhic_star,ref:rhic_phenix,ref:rhic_brahms,ref:rhic_phobos}. 
A large suppression of $\jpsi$ was proposed as a signature of the formation 
of QGP and was expected to arise 
from the color screening of the heavy quark potential in a deconfined medium~
\cite{ref:matsui}. 

Additional modifications of $\jpsi$ production due to cold nuclear matter (CNM
) effects~\cite{ref:RamonaNoEPS09} are expected. 
CNM effects are due to the presence of ordinary nuclear matter in the 
collision.
These include modifications to the Parton Distribution Functions (PDFs) 
inside a nucleus (shadowing, anti-shadowing, European Muon Collaboration (EMC) effect)~\cite{ref:shadowing,ref:emc_effect}  
and final-state nuclear absorption of $\jpsi$ by hadronic comovers~\cite{
ref:absorption}. 
In addition, the Cronin effect, which may be originating from multiple 
scattering of partons, should increase the mean $\pt$ of $\jpsi$ produced in $
\aaa$ collisions relative to $\pp$ collisions~\cite{ref:kopeliovich1, 
ref:kopeliovich2}.
In order to isolate the CNM effects and thereby improve our understanding of 
modifications to $\jpsi$ production in heavy-ion collisions, the production 
of $\jpsi$ is studied in both $\pp$ and $\dau$ collisions at Relativistic 
Heavy Ion Collider (RHIC), where the formation of a QGP was not originally 
expected.
Furthermore, $\jpsi$ production in $\pp$ collisions can provide information 
on the $\jpsi$ production mechanism in elementary collisions~\cite{
ref:starauau}. 

In this paper, the results for $\jpsi$ production at midrapidity ($|y|<1$, 
where $y$ is defined as $y=0.5\ln(\frac{E+p_{L}c}{E-p_{L}c})$, $E$ is the 
particle`s energy, $p_{L}$ is the particle`s momentum along beam axis and $c$ 
is the speed of light in a vacuum) 
in $\pp$ ($\pt < 4 \ \gevc$) and $\dau$ ($\pt < 3 \ \gevc$) collisions at $
\snn = 200 \ \gev$ from the STAR experiment are presented. The $\pp$ data 
were collected in 2009 and $\dau$ in 2008.
The $\pt$ spectrum from $\pp$ is combined with the high-$\pt$ STAR 
results~\cite{ref:starauau} and the resulting $\pt$ spectrum is compared with 
predictions from model calculations, including 
the color glass condensate (CGC) along with the non-relativistic quantum 
chromodynamics (QCD) based model with color singlet and color octet (CS+CO) 
contributions~\cite{ref:cgcnrqcd}, the NRQCD based CS+CO model only~\cite{
ref:nrqcd}
 and the color evaporation model (CEM)~\cite{ref:cem_star2}. 
The $\meanptt$ in both $\pp$ and $\dau$ collisions is calculated from the 
measured invariant cross sections and compared to measurements at other 
relevant collision energies. The value of $\meanptt$ is related to the width 
of the $\pt$ spectrum and is conventionally used to describe the Cronin effect
~\cite{ref:satz_pt2} in model calculations. It is described in more details 
further in the paper.

To quantify the CNM effects, the $\jpsi$ nuclear modification factor in $\dau$
 ($\rda$) has been calculated from the ratio of the invariant cross section 
in $\dau$ ($d^{2}\sigma_{\dau}/d\pt{d}y$) and $\pp$ (
$d^{2}\sigma_{pp}/d\pt{d}y$), 
    scaled by the average number of binary collisions $\mncoll$ according to 
the equation:

\begin{equation}
\rda = \frac{1}{\mncoll}\frac{d^{2}\sigma_{\dau}/d\pt
{d}y}{d^{2}\sigma_{pp}/d\pt{d}y}.
\label{eq:rda}
\end{equation}
The $\jpsi$ $\rda$ is compared to model predictions 
, which include cold nuclear matter effects and the modification of nuclear 
PDFs (nPDFs) using the EPS09~\cite{ref:EPS09} 
 and nDSg~\cite{ref:nDSg} parameterizations. 
Each of these models includes a final-state $\jpsi$ nuclear absorption cross 
section ($\sigabs$)~\cite{ref:RamonaNoEPS09, ref:RamonaEPS09} as an 
additional parameter, which can be determined by fitting the model 
calculations to the data.

The experimental setup and data used in this analysis are described in Section
~\ref{sec:Experiment} followed by a review of the analysis methods, $\jpsi$ 
signal reconstruction, 
    and corrections in Section~\ref{sec:Analysis}. The systematic 
uncertainties are explained in Section~\ref{sec:Systematics}, 
    and the results are described in Section~\ref{sec:Results}. Finally, a 
summary is provided in Section~\ref{sec:Summary}. 


\section{Experiment and Data\label{sec:Experiment}}
STAR~\cite{ref:star_det} is a large acceptance midrapidity (pseudorapidity $
|\eta|<1$ and full azimuthal angle) experiment at RHIC with excellent 
particle identification capabilities. It also includes additional detectors 
at forward and backward pseudorapidities ($|\eta|>1$) like the Forward Time 
Projection Chamber (FTPC) and Vertex Position Detector (VPD), and others, 
which are not used in this analysis.
The Time Projection Chamber (TPC) is the primary detector used for particle 
tracking and hadron identification while the Barrel Electro-Magnetic 
Calorimeter (BEMC) is used for electron identification.
For the $\pp$ data, the Time of Flight detector~(TOF)~\cite{ref:tof_det} was 
available for particle identification.
For the $\dau$ data the collision centrality was determined using the East 
Forward Time Projection Chamber (FTPC-E)~\cite{ref:ftpc_det}, which covers $-4
 < \eta < -2.5$.

The data used in this analysis were obtained from $\pp$ collisions recorded 
in 2009 and from $\dau$ collisions recorded in 2008 using a minimum bias (MB) 
trigger. 
The MB trigger was generated from the Vertex Position Detectors (VPDs)~\cite{
ref:vpd_det} to select collisions with a vertex position $|\vz| < 30 \ \cm$ 
by requiring coincidence signals within a bunch crossing in its East (gold 
facing in the case of $\dau$ collisions) and West detectors, both located $5.7 \ 
\mathrm{m}$ from the center of the TPC. 
The collision $\vz$ used in the trigger was evaluated from the time 
difference between VPD signals.
The $\dau$ trigger also required at least one neutron in the East Zero Degree 
Calorimeter (ZDC)~\cite{ref:zdc_det}, 
positioned $18 \ \mathrm{m}$ away from the center of the TPC. 

The offline vertex position was reconstructed using tracks in the TPC.
To remove out-of-time (pile-up) events in the $\pp$ data sample, a difference 
between the reconstructed and VPD vertex position $|\Delta\vz| < 6 \ \cm$ was 
required~\cite{ref:vpdref}. This removes $\sim15\%$ of events and leaves $\sim
 2\%$~\cite{ref:vpdref} of pile-up events. In $\dau$, pile-up events were 
removed by requiring at least one track from the collision be matched to the 
BEMC~\cite{ref:bemc_det}, which is a fast detector (readout time $\sim 10 \ 
\textrm{ns}$) and is not affected by pile-up. The BEMC match requirement 
along with $|\vz| < 30 \ \cm$ cut rejects $\sim 35\%$ of events with the 
possible bias estimated at $\sim 4\%$ at most.
The final data samples used in this analysis consisted of $7.7\times10^7$ $\pp
$ and $3\times10^7$ $\dau$ events satisfying the MB trigger and pile-up 
removal criteria.


\section{Analysis\label{sec:Analysis}}

\subsection{Collision geometry}

The collision centrality in $\dau$ reactions was determined using the Glauber 
model~\cite{ref:glauber} relating the measured particle multiplicity to the 
initial geometry.
The centrality selection in $\dau$ was obtained using the charged particle 
multiplicity in the FTPC-E~\cite{ref:ftpc_det} 
to minimize correlations between the centrality selection and the measured 
event in the TPC. 
The centrality definitions, the corresponding average number of participants (
$\mnpart$), number of collisions ($\mncoll$), 
and impact parameter ($\langle b \rangle$) in $\dau$ collisions are 
summarized in Table~\ref{tab:glauber}. 

A multiplicity dependent correction was performed in $\dau$ to account for 
the trigger bias towards events with high multiplicity. 
This was done by comparing the multiplicity distribution 
measured using the TPC and FTPC-E to the distributions obtained from the 
Glauber model to calculate a multiplicity dependent weight. The corrections were later applied using event-by-event reweighing, which increased the overall (event integrated) weight of events in the $40-100
\%$ centrality bin in 
$\dau$ collisions by a factor $\sim 1.33$, while having a negligible effect on 
semi-central and central collisions. The event integrated weights for $\dau$ are listed in Table~\ref{tab:mult_weights}. 
An overall trigger correction of $70\%$~\cite{ref:vpdref} has been used in MB 
$\pp$ collisions to account for the trigger bias towards events containing a $
\jpsi$ as discussed later.

\begin{table}[htbp]
\begin{center}
\caption{Event integrated weights used for trigger bias correction in $\dau$ in each centrality bin.}
\label{tab:mult_weights}
\begin {ruledtabular}
\begin {tabular} { c c }
\noalign{\smallskip}
Centrality & Weight [1]\\ 
\noalign{\smallskip}\hline\noalign{\smallskip}
$0 - 40 \%$  & $1$	\\ 
$40 - 100\% $ & $\sim 1.33$\\ 
\end {tabular}
\end {ruledtabular}
\end{center}
\end{table}

\begin{table}[htbp]
\begin{center}
\caption{The collision centrality and geometry definitions from the Glauber 
model in $\dau$ collisions ~\cite{ref:star_dau_centrality}. The listed errors 
are systematic only.}
\label{tab:glauber}
\begin {ruledtabular}
\begin {tabular} { c c c c }
\noalign{\smallskip}
Centrality & $\mnpart$ & $\mncoll$ & $\langle b \rangle \ (\fm)$\\ 
\noalign{\smallskip}\hline\noalign{\smallskip}
$0 - 40 \%$  & $13.3 \pm 2.3$	& $12.7 \pm 2.3$  & $4.1 \pm 0.8$	\\ 
$40 - 100\% $ & $5.7 \pm 0.7$   & $4.8 \pm 0.6$	  & $6.7 \pm 0.8$	\\ 
\end {tabular}
\end {ruledtabular}
\end{center}
\end{table}

\subsection{Particle identification}

\begin{figure}[h]
\begin{center}

$\begin{array}{c}
(a)\\
\includegraphics[width=0.44\textwidth]{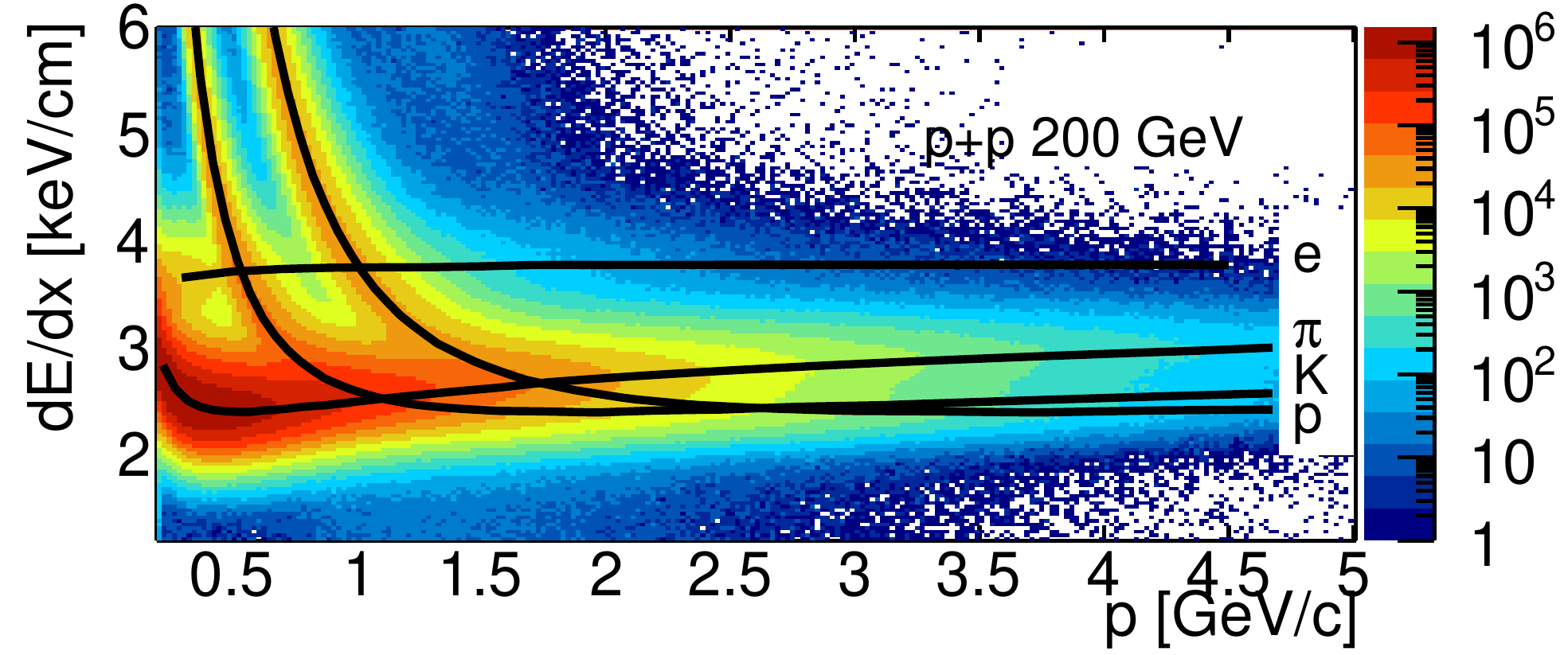}
\\
(b)\\
\includegraphics[width=0.44\textwidth]{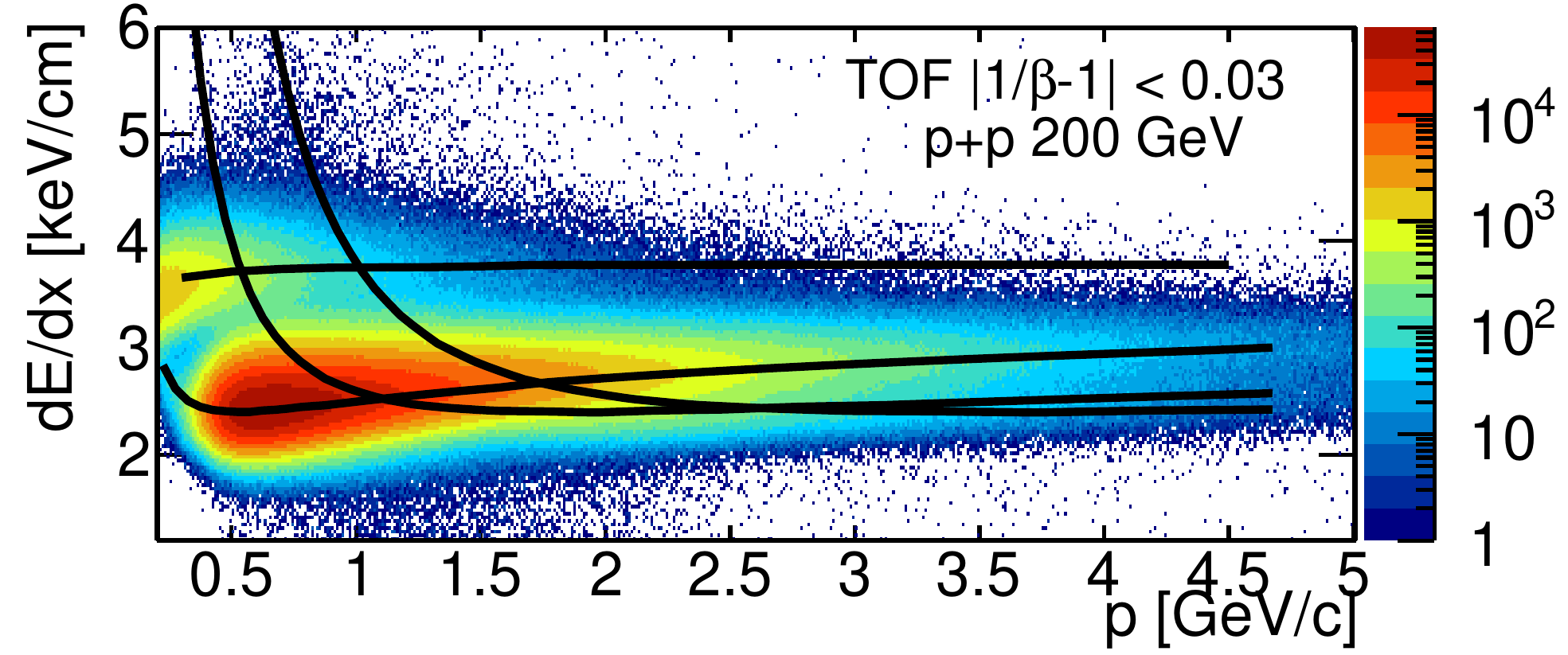}
\\
(c)\\
\includegraphics[width=0.44\textwidth]{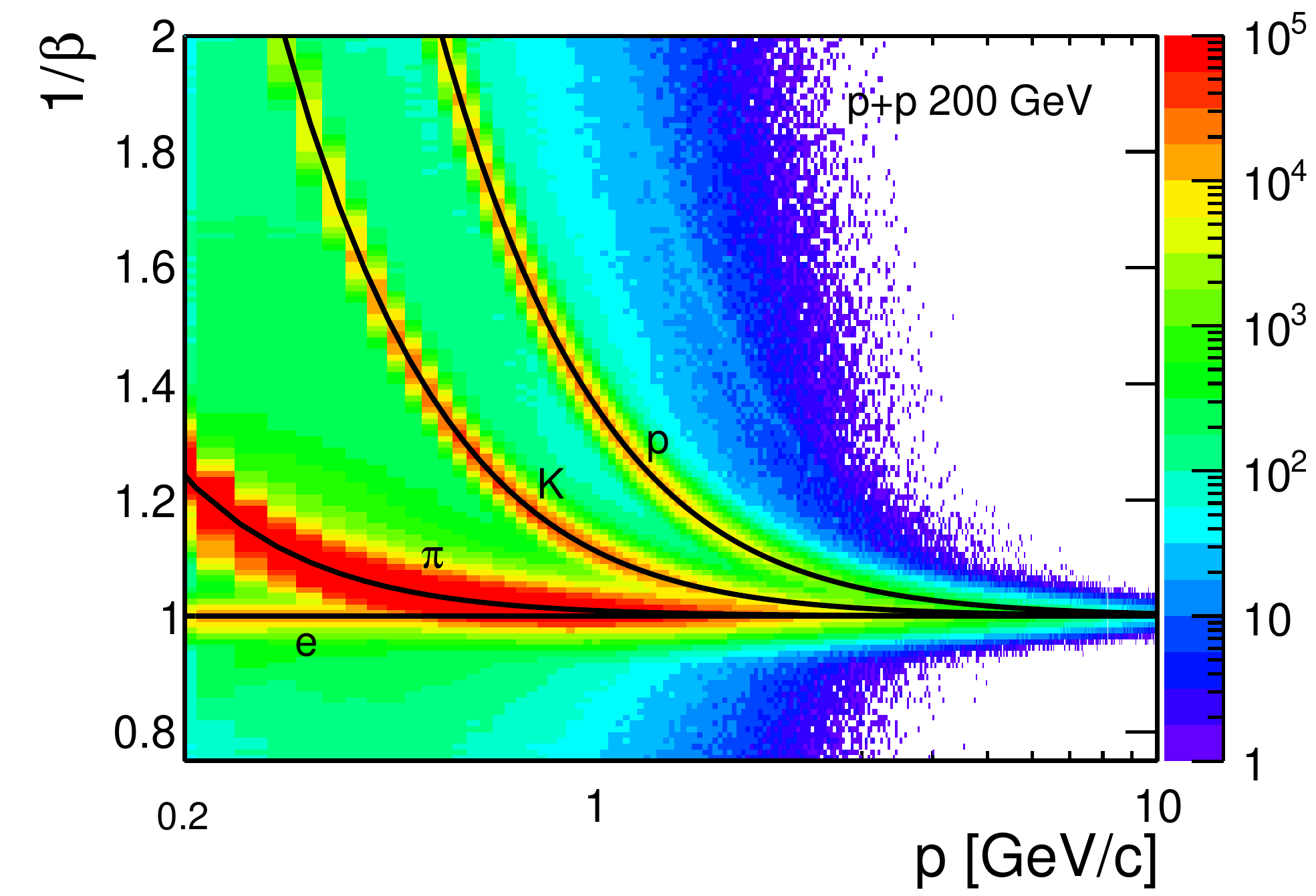}
\\
\end{array}$

\caption{
  (Color online) (a) The ionization energy loss $\dedx$ versus momentum for 
charged particles in $\pp$ collisions. The lines indicate the expected $\dedx$
 for various particles 
    obtained from the Bichsel functions~\cite{ref:bichsel}. (b) The $\dedx$ 
distribution after removing slower hadrons using the TOF. (c) The TOF $
\invbeta$ versus momentum for charged particles in $\pp$ collisions. The 
lines indicate the expected values for various particles. 
}
\label{fig:dedx}
\end{center}
\end{figure}

The reconstruction of $\jpsi$ has been performed using the dielectron decay 
channel $\jpsi \rightarrow e^{+}+e^{-}$ with a branching ratio of $B_{ee} = 5.
961\pm0.033\%$ ~\cite{ref:pdgref}. 
Electrons and positrons were identified from the ionization energy they 
deposited in the TPC.
The $\dedx$ versus momentum for charged particles in the TPC is shown in Fig.~
\ref{fig:dedx}(a). The lines indicate the expected $\dedx$ for various particles 
obtained from the Bichsel functions~\cite{ref:bichsel}. 

\begin{figure}[h]
\begin{center}
$\begin{array}{c}
\includegraphics[width=.5\textwidth]{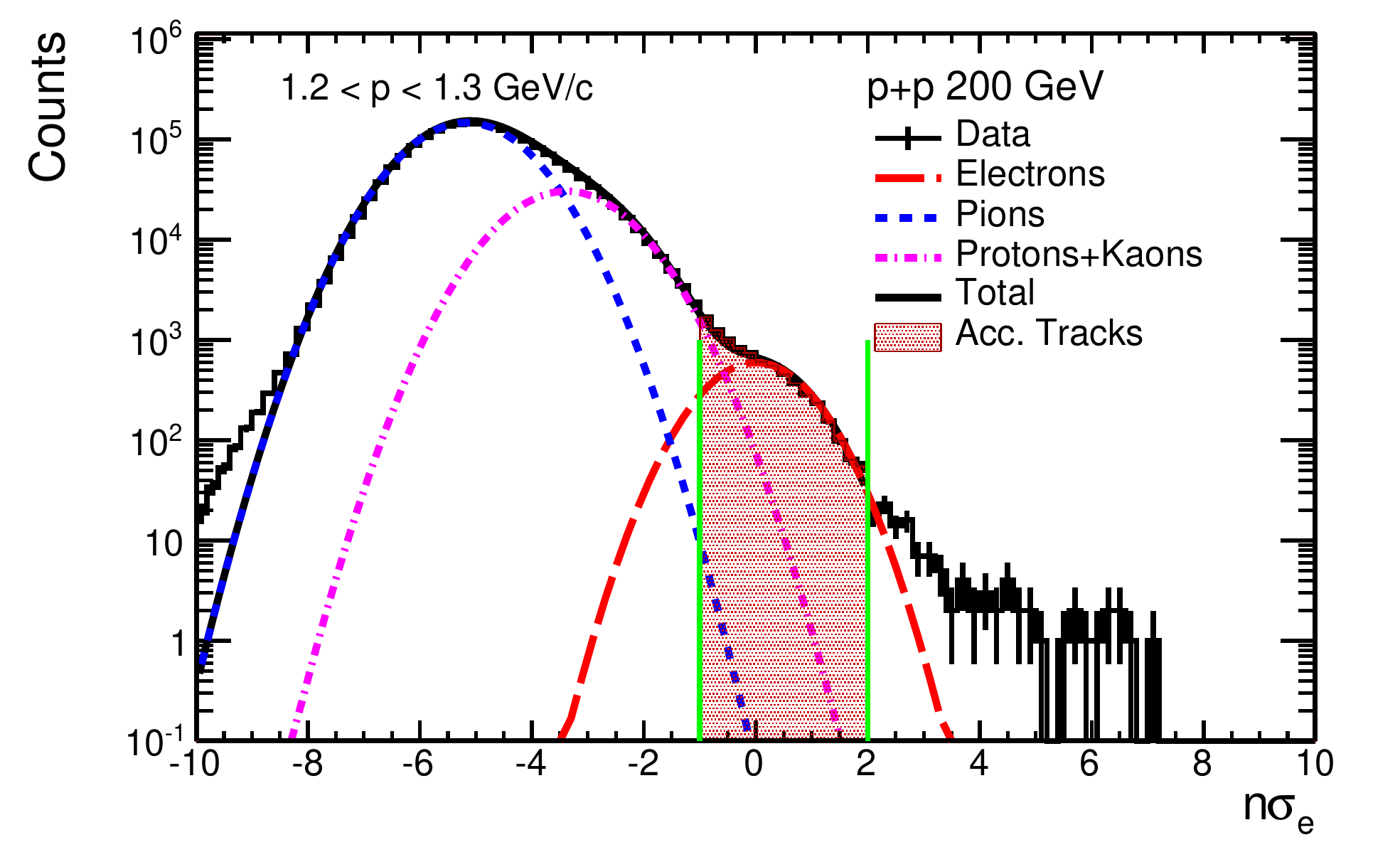} \\
\includegraphics[width=.5\textwidth]{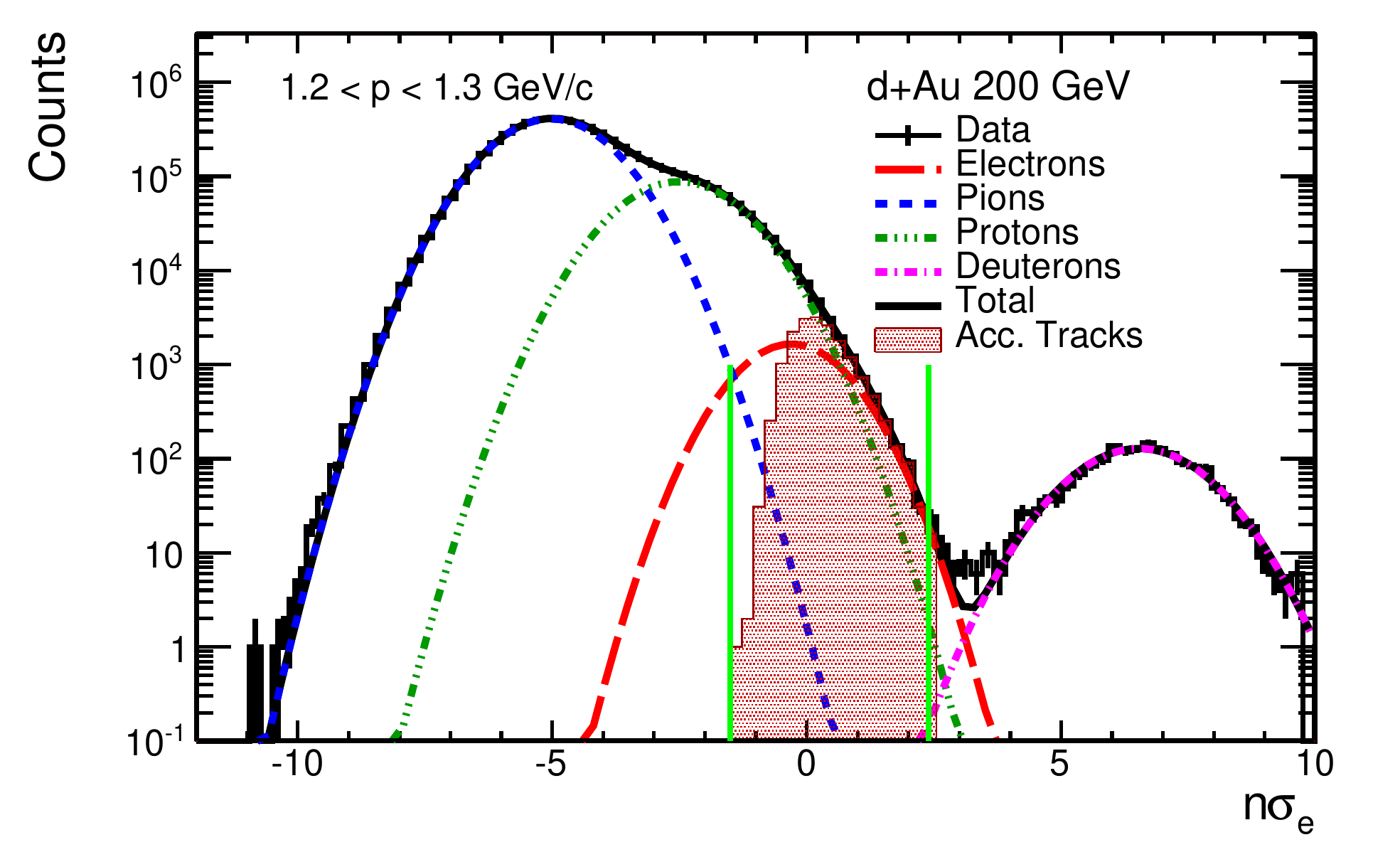} 
\end{array}$
\caption{
  (Color online) The $\dedx$ $\nsig$ distributions for all charged particles 
in $\pp$ (upper plot) and $\dau$ (lower plot) collisions with  $1.2 < p < 1.3 
\ \gevc$. Multiple Gaussians have been fitted to the particle distributions. 
The vertical lines indicate the accepted range, and the shaded region 
	shows the accepted particles. The plots are after TOF cut for $\pp$ and 
BEMC cuts for $\dau$.
}
\label{fig:nsigma}
\end{center}
\end{figure}

The deviation of the measured $\dedx$ from the expected $\dedx$ for an 
electron, $\nsig$, is defined here as:
\begin{equation}
\nsigal = \ln\left(\frac{\dedx|_{Measured}}{\dedx|_{Expected}}\right)/\sigma, 
\: \alpha = e,\pi,K,p.
\end{equation}
where $\dedx|_{Measured}$ is measured with the TPC, $\dedx|_{Expected}$ is 
the expected value and $\sigma$ is the resolution of the measured $\ln(\dedx)$
. 
To remove the large hadron contamination at low momenta where the $\dedx$ of 
electrons and hadrons overlaps, a minimum transverse momentum ($\pt$) 
of electron candidates was applied. Only tracks with $\pt > 0.8 \ \gevc$ in $
\pp$ and $\pt > 1 \ \gevc$ in $\dau$ were accepted. 
The $\nsig$ distribution for charged particles with $1.2 < p < 1.3 \ \gevc$ 
in $\pp$ and $\dau$ collisions is shown in Fig.~\ref{fig:nsigma} 
and has been fitted with the sum of Gaussian distributions to account for the 
individual particle contributions. 
The vertical lines indicate the accepted range, and the shaded region 
represents the electron candidates.
Electrons were selected by requiring $-1 < \nsig < 2$ in $\pp$ and $|\nsig|<2.
4$ in $\dau$. The asymmetric cut in $\pp$ was used to remove the large hadron 
contamination 
at $\nsig < -1$. These hadrons in $\dau$ were rejected by requiring $|\nsigP| 
> 2.2$ and $|\nsigPi| > 2.5$. 
These cuts lead to the non-regular shape of the left side of shaded area in 
the  Fig.~\ref{fig:nsigma} (lower panel), where the horizontal scale is in $\nsig$
 units.
The hadron $\dedx$ rejection cuts were not necessary in the $\pp$ analysis, 
as the TOF was used to separate electrons from slow hadrons and also allowed 
for a lower cut on the minimum $\pt$ in $\pp$ collisions. 
Because the TOF detector is only available in the $\pp$ sample, the particle 
identification cuts are different in $\pp$ and $\dau$ analyses. The 
differences are summarized in Table~\ref{tab:cuts}.

\begin{table}[htbp]
\begin{center}
\caption{Summary of analysis cuts in $\pp$ and $\dau$.}
\label{tab:cuts}
\begin {ruledtabular}
\begin {tabular} { c c c }
\noalign{\smallskip}
Cut & $\pp$ & $\dau$\\ 
\noalign{\smallskip}\hline\noalign{\smallskip}
$\pt$ & $> 0.8 \ \gevc$  & $> 1 \ \gevc$ \\ 
$\nsig$ & $-1 < \nsig < 2$ & $< 2.4$ \\ 
$|\nsigP|$ & $-$ & $> 2.2$ \\ 
$|\nsigPi|$ & $-$ & $> 2.5$ \\ 
$|\invbeta-1|$ & $< 0.03$ for $p < 1.4 \ \gevc$ & $-$ \\ 
$\eop$ & $> 0.5$ for $p > 2 \ \gevc$ & $> 0.5$ \\ 
\end {tabular}
\end {ruledtabular}
\end{center}
\end{table}

In the $\pp$ data sample, each particle\textquotesingle s velocity ($\beta$) 
is evaluated using the TOF detector.
The TOF $\invbeta$ distribution in $\pp$ collisions 
is shown in Fig.~\ref{fig:dedx}(c) versus the momentum obtained from the track 
reconstruction in the TPC. 
The lines indicate the expected $\invbeta$ values for several particle 
species. The 
$72\%$ of the TOF detector was installed for the $\pp$ data and it was used 
to improve the 
electron identification for $p < 1.4 \ \gevc$. At higher momenta, the $
\invbeta$ of electrons and hadrons converges to unity and the electron-hadron 
discrimination power decreases. 
Heavier hadrons were removed by requiring $\invbetacut$. The resulting $\dedx$
 distribution is shown in Fig.~\ref{fig:dedx}(b). 
This cut successfully removes most of the contributions from kaons, protons, 
and deuterons. The remaining pions are sufficiently well separated from the 
electrons and were removed using the TPC $\dedx$. 

\begin{figure}[h]
\begin{center}
\includegraphics[width=.5\textwidth]{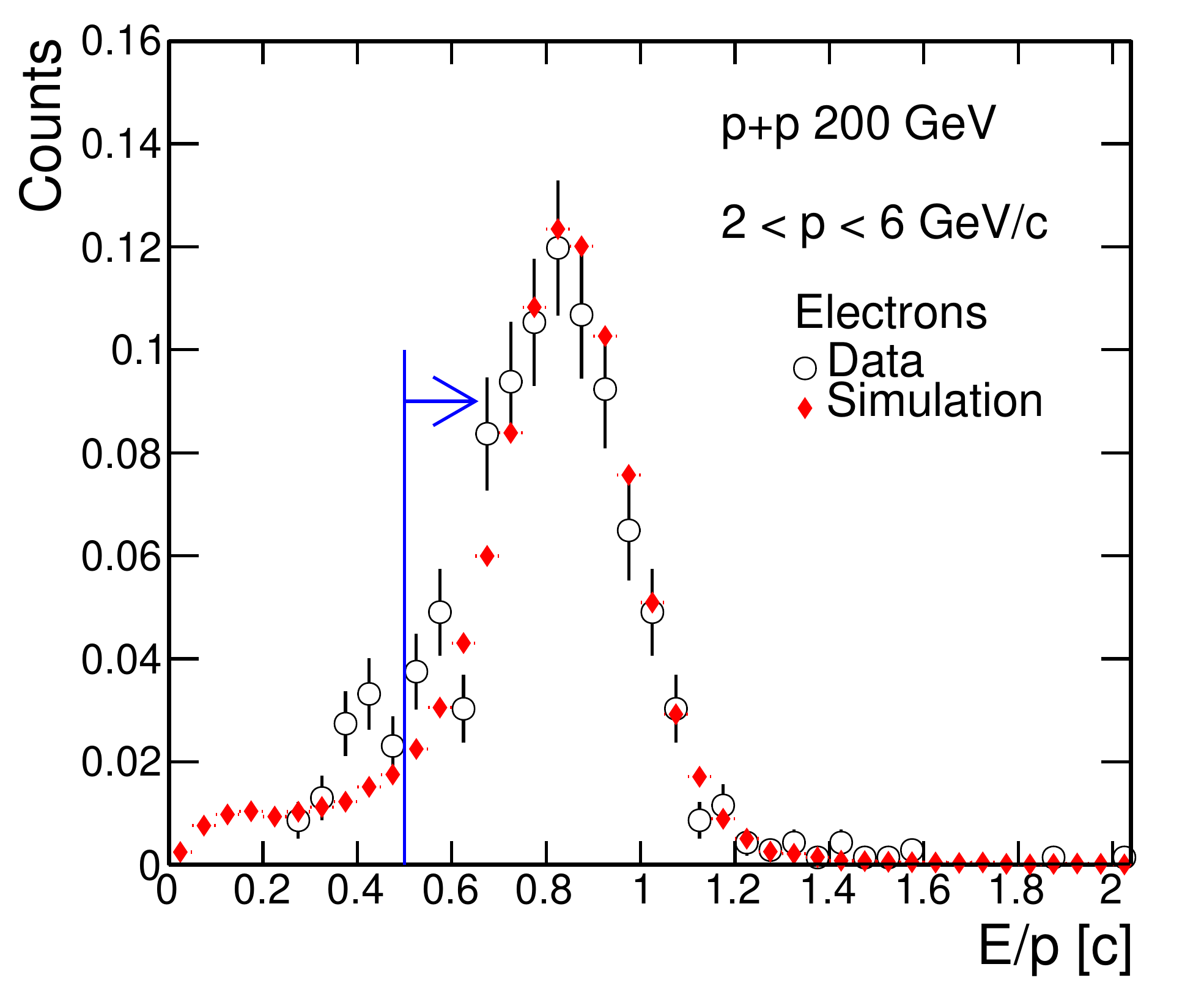} 
\caption{
  (Color online) The $\eop$ distribution for electrons with $2 < p < 6 \ \gevc
$ in $\pp$ collisions from data (closed circles) and simulation (closed 
diamonds).  
}
\label{fig:eop}
\end{center}
\end{figure}

Electron energy is measured using the BEMC.
The BEMC has a radiation length of $20X_{0}$ and is segmented into towers of 
dimension $\Delta\eta\times\Delta\phi = 0.05 \times 0.05$. 
The energy deposited in its towers has been used to calculate the energy-to-
momentum ratio ($\eop$), which should be $\sim 1$ for electrons. 
The $\eop$ distribution is shown in Fig.~\ref{fig:eop} for electrons obtained 
from data and a Monte Carlo GEANT~\cite{ref:glauber} simulation with STAR 
detector geometry.
A non-Gaussian tail at low-$\eop$ results from electrons striking near the 
edge of the tower and sharing their energy between multiple towers. This also 
causes a shift of the Gaussian to low $\eop$ values.
Electrons are identified and selected by requiring $\eop > 0.5$ for $p > 2 \ 
\gevc$ in $\pp$ and $\eop > 0.5$ in $\dau$ collisions. 

\subsection{\bjpsi signal}

\begin{figure}[h]
\begin{center}
\includegraphics[width=.5\textwidth]{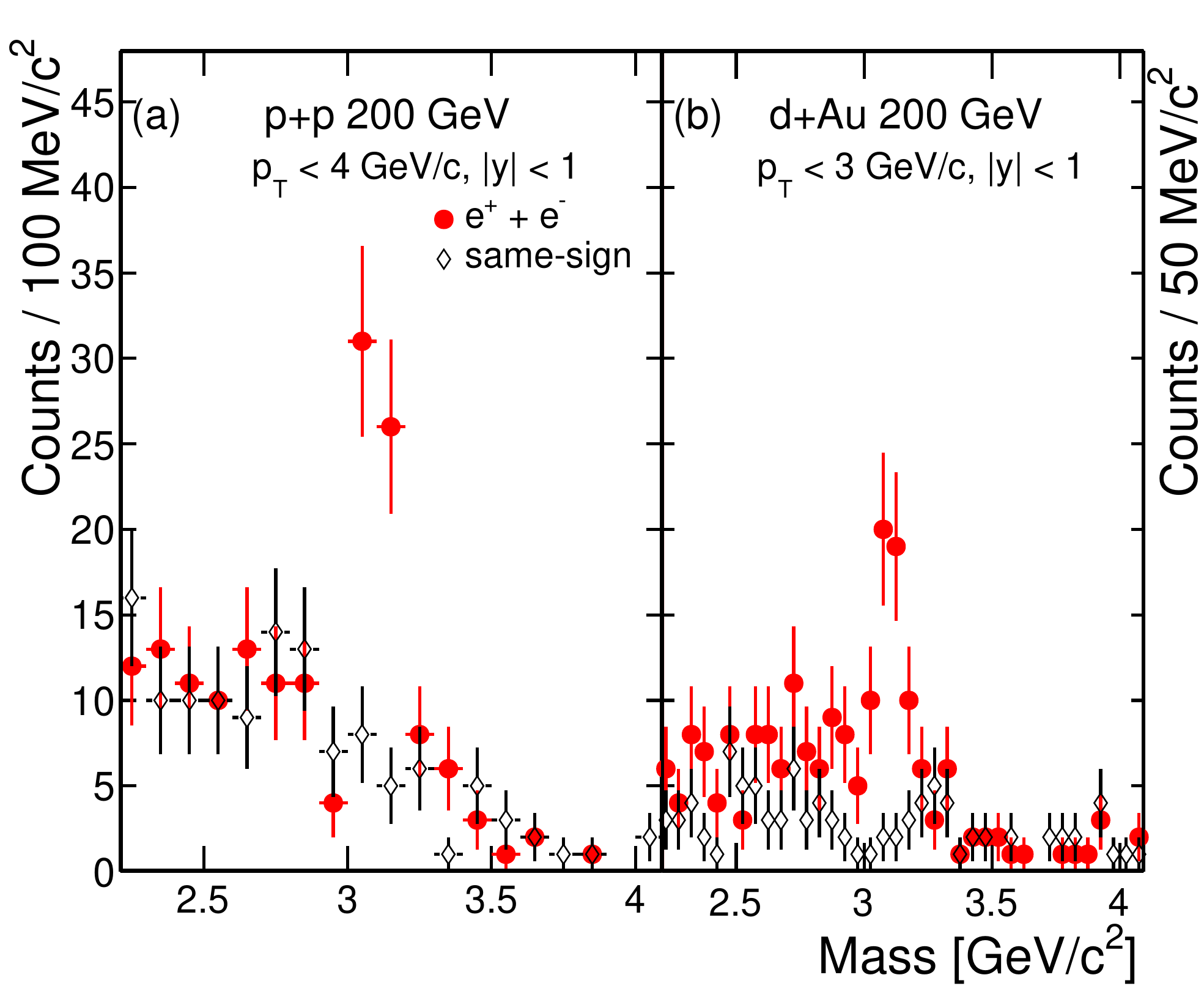}  
\caption{
  (Color online) The opposite-sign dielectron invariant mass distribution in (
a) $\pp$ collisions and (b) $\dau$ collisions at $\snn = 200 \ \gev$ (closed 
circles). 
    The combinatorial background (open diamonds) has been calculated from the 
same-sign pairs. 
}
\label{fig:mass}
\end{center}
\end{figure}

The dielectron invariant mass spectrum is constructed from electrons 
identified using TPC, BEMC, and TOF.
The resulting dielectron invariant mass spectra in $|y|<1$ for $\pt < 4 \ 
\gevc$ in $\pp$ and 
$\pt < 3 \ \gevc$ in 
$\dau$ collisions at $\snn = 200 \ \gev$ are shown in Fig.~\ref{fig:mass}. 
The combinatorial background has been calculated using a sum of same-sign 
electron pairs ($e^{+}e^{+}+e^{-}e^{-}$), and a 
signal-to-background ratio of $S/B = 0.81$ in $\pp$ and $S/B = 2.3$ in $\dau$ 
was obtained for $2.7 < m < 3.2 \ \gevcc$. 

\begin{figure}[h]
\begin{center}
\includegraphics[width=.5\textwidth]{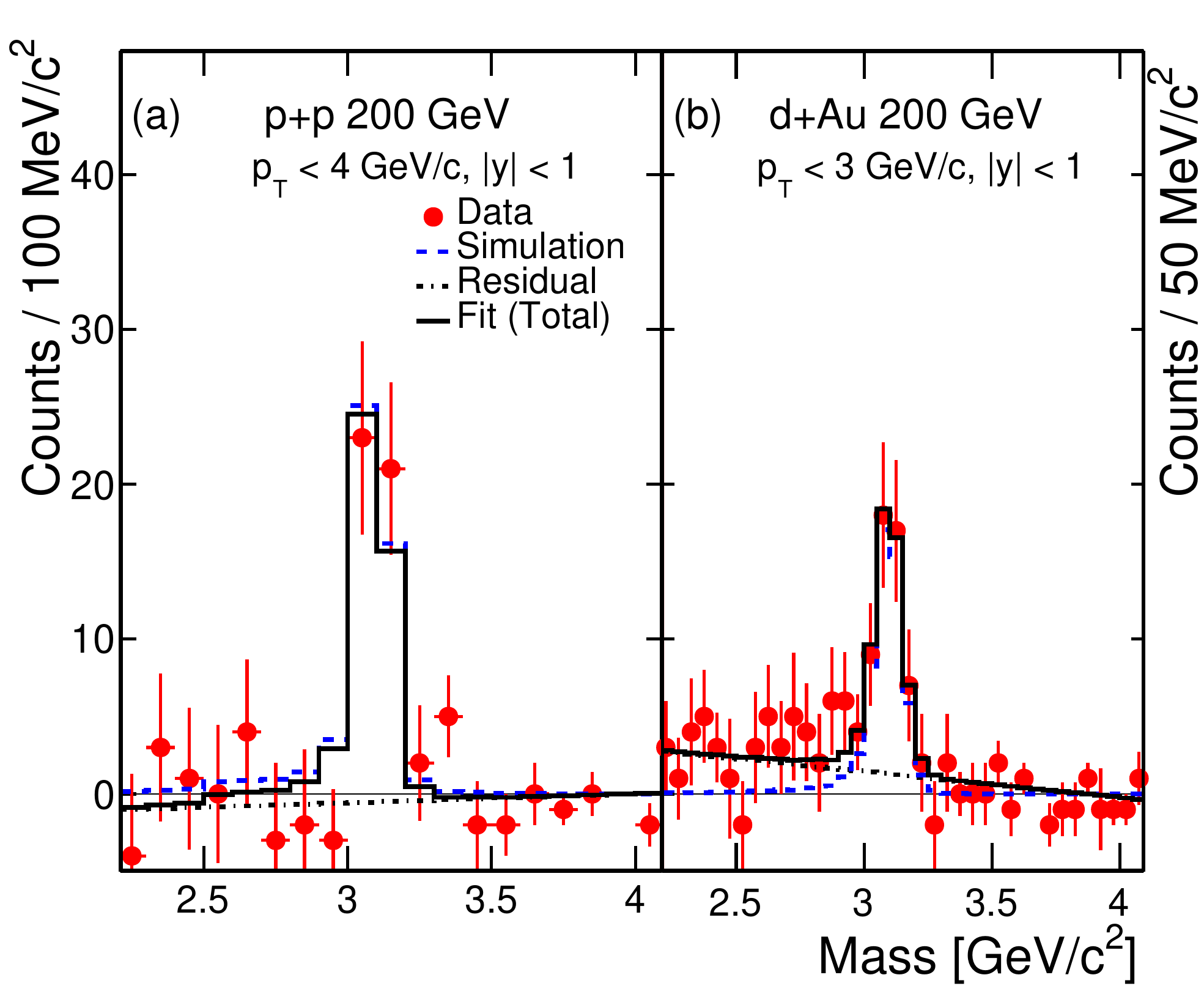}  
\caption{
  (Color online) The $\jpsi$ signal for $|y|<1$ in (a) $\pp$ collisions and (b
) $\dau$ collisions at $\snn = 200 \ \gev$, after same-sign background 
subtraction (closed circles).  
    The signal shape obtained from simulation (dashed line) is combined with 
a residual background (dot-dashed line), and the total is fitted to 
    the data (solid line). 
}
\label{fig:signal}
\end{center}
\end{figure}

The $\jpsi$ signal obtained from subtracting the combinatorial background 
from the dielectron invariant mass spectrum is shown in Fig.~\ref{fig:signal}. 
The signal shape has been obtained from a Monte Carlo GEANT simulation 
and reflects the TPC momentum resolution and energy loss in the detector 
material. 
This shape is combined with a straight line to account for a residual 
background ($c\bar{c}$ continuum, Drell-Yan), and the total has been fitted 
to the data. The yield has been calculated by performing a bin counting of 
the data entries in the range $2.7 < m < 3.2 \ \gevcc$ in $\pp$ and $\dau$ 
collisions. The residual background has been subtracted, and the counts have 
been corrected for the number of $\jpsi$ outside of this mass range using the 
signal shape from simulation. 
A total of $44\pm 14$ ($66\pm 10$) $\jpsi$s were obtained in $\pp$ ($\dau$) 
collisions with a significance of $3.2\sigma$ ($6.8\sigma$). 

\begin{figure}[h]
\begin{center}
\includegraphics[width=.5\textwidth]{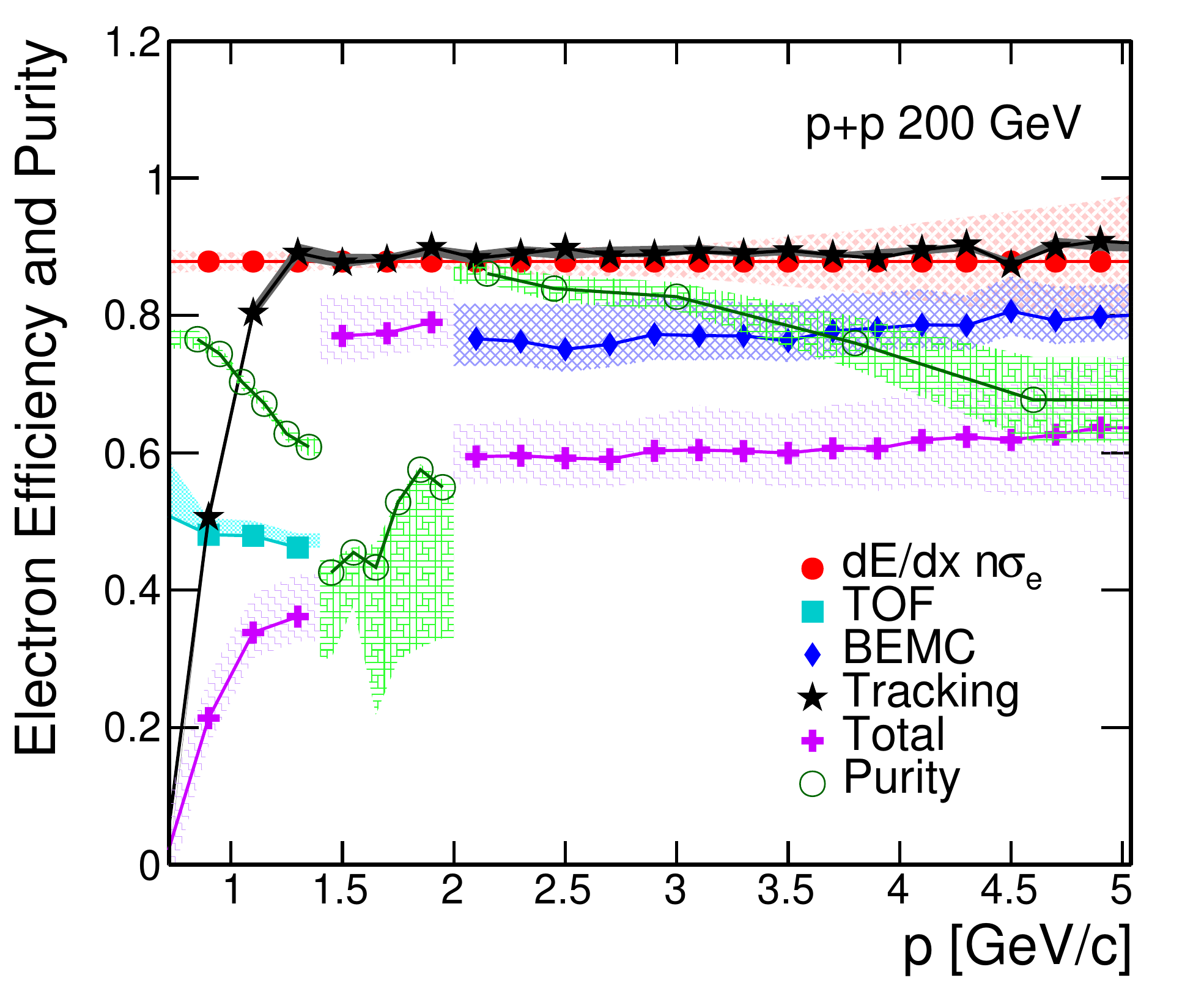} 
\caption{
  (Color online) The single electron efficiencies, including the $\dedx$ (
closed circles), TOF (closed squares), BEMC (closed diamonds), tracking (
closed stars), 
    and total (closed crosses) efficiencies and purity (open circles) in $\pp$
 collisions. The shaded regions represent systematic uncertainties.
}
\label{fig:efficiency_pp}
\end{center}
\end{figure}

\subsection{Corrections}

The electron identification efficiency is defined as the ratio of accepted 
electrons to all electrons, while purity is the fraction of electrons in the 
selected sample. 
This is illustrated in Fig.~\ref{fig:nsigma}, where the electron contribution 
of the $\nsig$ distribution in $\pp$ collisions has been fit with a Gaussian 
function. 
The vertical lines indicate the accepted range, and the shaded region 
indicates the accepted particles. 
The efficiency is the integral of the electron Gaussian within the green 
lines ($-1 < \nsig < 2$) divided by the integral over the entire electron 
Gaussian (dashed red line).
The purity is calculated by taking the ratio of the integral of the electron 
Gaussian function within the green lines to the integral of the accepted 
tracks histogram (shaded area).

The efficiencies related to the electron identificiation requirements in $\pp$
 are shown in Fig.~\ref{fig:efficiency_pp}. 
The $\pp$ analysis uses different detectors to identify particles at 
different momenta, due to their different performance, as was explained in 
the previous section. 
Electron identification is determined from the TOF and TPC at low momentum ($
p < 1.4 \ \gevc$), the BEMC and TPC at high momentum ($p > 2 \ \gevc$), and 
the TPC alone for $1.4 < p < 2 \ \gevc$.
The abrupt changes in Fig.~\ref{fig:efficiency_pp} arise from the different 
efficiencies of each detector. 
The $\dedx$ cut efficiency is mostly constant 
as no requirements were placed on the hadrons. The BEMC was used for $p > 2 \ 
\gevc$ and the combined matching and $\eop$ efficiency are $\sim 80\%$, 
   consistent with the BEMC efficiency in $\dau$. The TOF has been used for $
p < 1.4 \ \gevc$, and the matching and $\invbeta$ efficiency are combined 
with the 
   TOF accceptance to obtain a correction  of $\sim 50\%$. The electron 
tracking efficiency is $\sim 90\%$ for $p > 1.4 \ \gevc$, and decreases below 
this due to the 
   minimum $\pt$ required for electrons. The total efficiency and purity are 
also shown, and a sudden drop is observed for $1.4 < p < 2 \ \gevc$ where 
only the TPC is used for particle identification in order to maximize 
statistics.

\begin{figure}[hbt!!!!!!!!!!!!]
\begin{center}
\includegraphics[width=.5\textwidth]{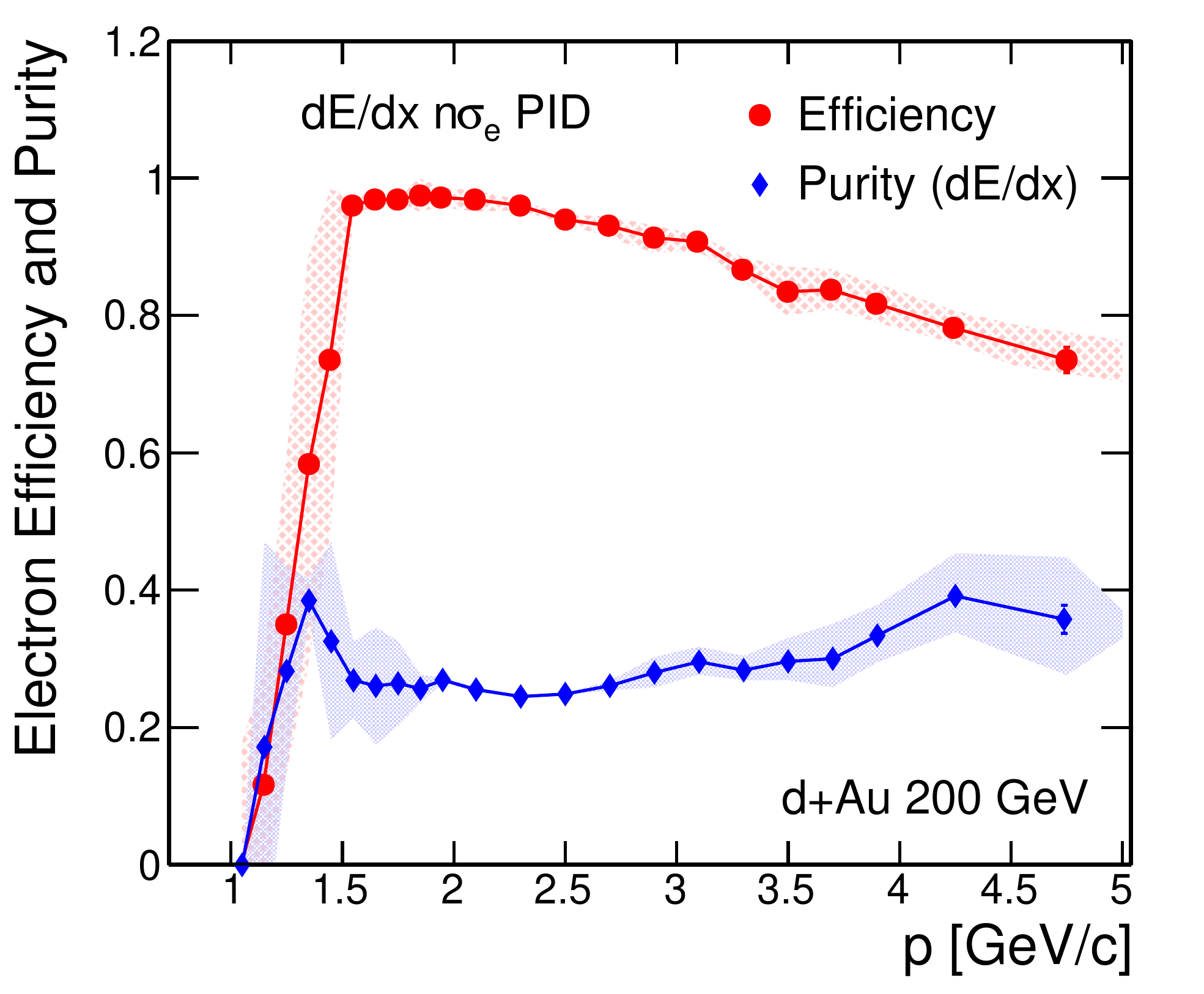} 
\caption{
  (Color online) The single electron $\dedx$ identification efficiency (
circles) and purity obtained in $\dau$ collisions. The shaded areas 
correspond to systematic uncertainty.
}
\label{fig:efficiencyDedx}
\end{center}
\end{figure}

The efficiency associated with the $\dedx$ electron identification 
requirements in $\dau$ collisions is shown in Fig.~\ref{fig:efficiencyDedx}. 
Electron identification in $\dau$ was performed using the TPC $\dedx$ and 
BEMC $\eop$ for $\pt > 1 \ \gevc$. 
The $\dedx$ identification efficiency in $\dau$ for $p < 1.4 \ \gevc$ is 
smaller than the efficiency in $\pp$ due to the hadron rejection cuts as 
discussed earlier. At high $\pt$, 
the efficiency decreases slightly due to the relativistic rise of the hadron $
\dedx$ (see Fig.~\ref{fig:dedx}a). 
The purity of the electron sample obtained using the TPC $\dedx$ alone is 
depicted in Fig.~\ref{fig:efficiencyDedx}. The purity increases by a factor $
\sim 2$ when including the $\eop$ cut.

\begin{figure}[hbt!!!!!!!!!!!!!]
\begin{center}
\includegraphics[width=.5\textwidth]{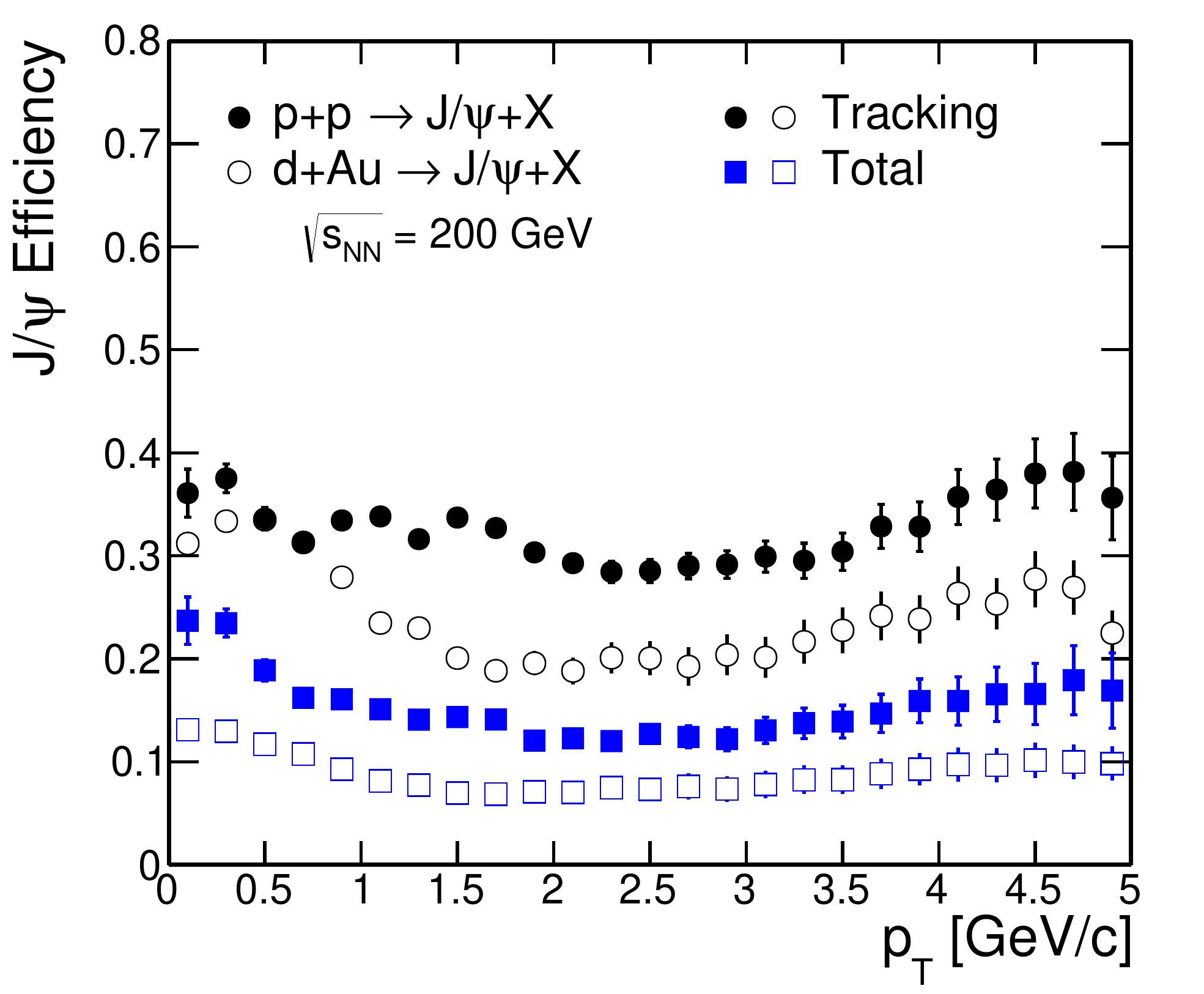} 
\caption{
  (Color online) The $\jpsi$ tracking efficiency and acceptance (circles) is 
combined with the electron identification efficiency to determine the total $
\jpsi$ efficiency (squares) 
    in $\pp$ collisions (closed symbols) and $\dau$ collisions (open symbols)
.  
}
\label{fig:efficiencyTotal}
\end{center}
\end{figure}

The total $\jpsi$ tracking efficiency and acceptance in $|y|<1$ have been 
obtained from a GEANT simulation, and are shown in 
Fig.~\ref{fig:efficiencyTotal} for $\pp$ and $\dau$. A lower efficiency was 
observed in 
$\dau$ due to the higher minimum $\pt$ required for electrons. The tracking 
efficiency and acceptance 
have been combined with the electron identification efficiencies to obtain 
the total $\jpsi$ efficiency corrections also shown in 
Fig.~\ref{fig:efficiencyTotal}. 
An additional trigger correction of $70\%$~\cite{ref:vpdref} has been applied 
to the $\pp$ data to account for the VPD selection bias towards events 
containing a $\jpsi$. 
This $70\%$ correction was determined by comparing the number of events 
containing at least one $\jpsi$ to the number of unbiased events 
in a Monte Carlo PYTHIA simulation coupled with the STAR detector geometry~
\cite{ref:vpdref}.



\section{Systematic Uncertainties\label{sec:Systematics}}

The main sources of systematic uncertainties on the yield in $\pp$ and $\dau$ 
collisions arise from the uncertainty in the corrections and yield extraction 
procedure. 
These are investigated separately in each $\jpsi$ $\pt$ bin, while the 
uncertainties on the integrated yield are reported in the text.
The multiple Gaussian fits to the $\nsig$ distribution and the calculation of 
the efficiency from the $\dedx$ requirements using these fits resulted in an 
uncertainty of $6\%$ in $\pp$ and a higher $16\%$ in $\dau$ collisions due to 
lack of TOF.
The uncertainty on the BEMC matching and $\eop$ efficiency was found to be 
$9\%$ ($11\%$) in $\pp$ ($\dau$) and was estimated by comparing the 
efficiency obtained from a high purity electron sample from the data to the 
efficiency obtained from simulation.
An additional $4\%$ systematic uncertainty due to the TOF requirement in $\pp$
 collisions was estimated by comparing the efficiency from electrons and 
hadrons.
The tracking efficiency was obtained from a GEANT simulation, from which 
comparison 
between the track properties from simulation and data resulted in an 
uncertainty of $3\%$ in $\pp$ collisions. A higher uncertainty of $ 12\%$ in $
\dau$ collisions was obtained due to higher backgrounds.
The shape of the input rapidity and $\pt$ distributions in simulation 
were varied to determine the effect on the $\pt$-dependent and $\pt$-
integrated efficiency calculation, and an uncertainty on the final yield of $8
\%$ ($9\%$)
was determined from the efficiency correction in $\pp$ ($\dau$) collisions. 
The uncertainty on the yield was calculated by changing the mass window in 
which the counting was performed, and comparing 
the yield to that obtained from the integral of the signal shape from 
simulation.  
This resulted in an uncertainty of $40\%$ in $\pp$ collisions and $15\%$ in $
\dau$ collisions. 
An additional $4\%$ uncertainty due to the contribution from internal 
radiation ($\jpsi\rightarrow e^{+}e^{-}\gamma$) was also included. 
The effect of possible bias due to pile-up events removal with the BEMC in $
\dau$, was estimated to be $4\%$ at most.

The normalization uncertainty on the nuclear modification factor $\rda$ [Eq.~
(\ref{eq:rda})] combines the uncertainty on $\ncoll$ of $12\%$, 
    and the statistical and systematic uncertainty of the $\jpsi$ cross 
section in $\pp$ for $\pt < 3 \ \gevc$.  
The systematic uncertainty includes the normalization uncertainty for the 
inelastic cross section in $\pp$ ($\sigma^{\pp}$) 
  of $8\%$~\cite{ref:norm_uncertain}. 
The systematic uncertainties in $\pp$ for $\pt < 4 \ \gevc$ and $\dau$ 
collisions are summarized in Table~\ref{tab:systematics}.

\begin {table} [tbp]
\begin {center}
\caption{The systematic uncertainties on the yield in $\pp$ and $\dau$ 
collisions. } 
\label{tab:systematics}
\begin{tabularx}{0.48\textwidth}{X X X}
\noalign{\smallskip}
\hline\hline
\noalign{\smallskip}
\multirow{2}{*}{Source} & \multicolumn{2}{l} {Relative uncertainty (\%)} \\
                        & $\pp$ & $\dau$ \\
\hline\noalign{\smallskip}

eID  (TPC) & ${\pm 6}$ & ${+16} \ {-11}$ \\
eID (BEMC) & ${\pm 9}$ & $\pm11$ \\
eID (TOF) & ${\pm 4}$ & $-$ \\
Tracking & ${\pm 3}$ & $\pm 12$\\
Efficiency Corr.& ${\pm 8}$ & $\pm9$ \\
Yield & ${\pm 44}$ & ${\pm23}$ \\
Total & ${\pm 46}$ & ${\pm33}$ \\
\hline\noalign{\smallskip}
$\ncoll$ & $-$ & $\pm12$ \\
$\sigma^{\pp}$ & {$\pm 8$} & {$\pm 8$}\\

\noalign{\smallskip}
\hline\hline

\end{tabularx}
\end {center}
\end {table}


\section{Results\label{sec:Results}}

\begin{figure}[h] \begin{center} $\begin{array}{cc} (a) & (b) \\ 
\includegraphics[width=.24\textwidth]{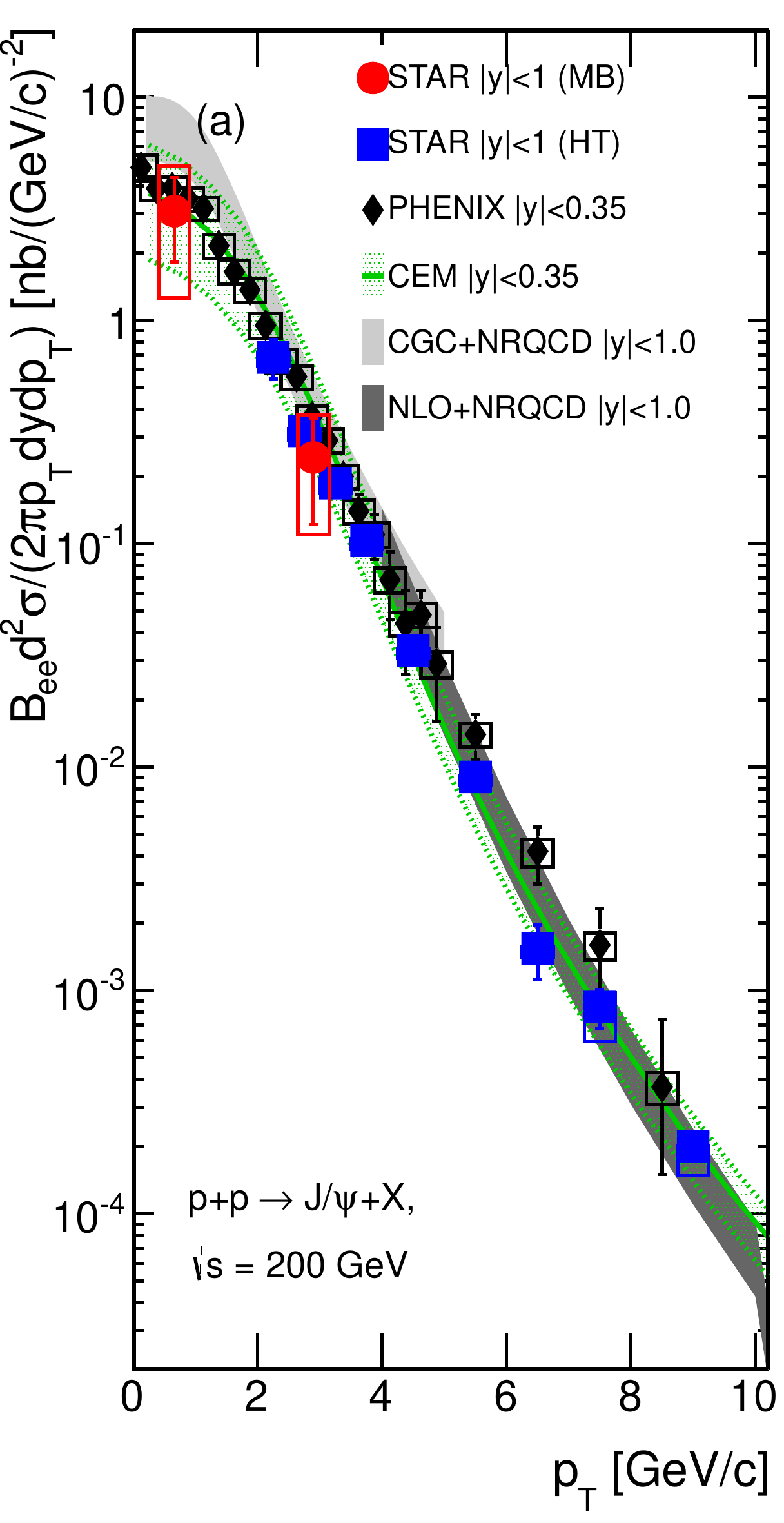} & \includegraphics[width=.24
\textwidth]{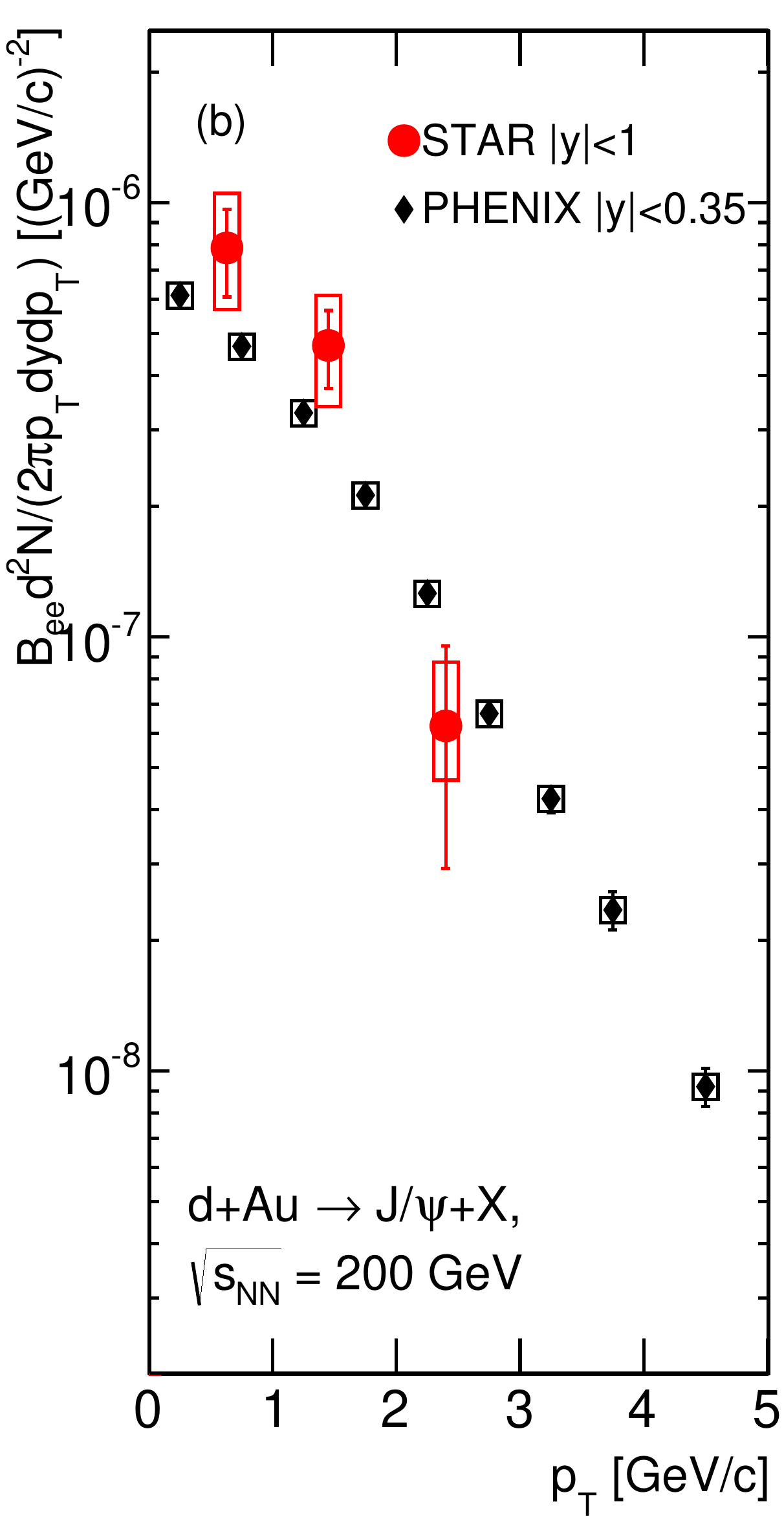}  \end{array}$ \caption{   (Color online) (a) The 
invariant cross section versus transverse momentum for $\jpsi$ with $|y|<1$ 
in $\pp$ collisions (closed circles), compared to high-$\pt$ STAR data~
\cite{ref:starauau} in $|y|<1$ (closed squares), PHENIX data in $|y|<0.35$~
\cite{ref:phenixpp_pol} (closed diamonds), and various model predictions~
\cite{ref:cem_star2, ref:cgcnrqcd, ref:nrqcd}. 
    (b) The invariant yield versus transverse momentum for $\jpsi$ with $|y|<1
$ in $0-100\%$ central $\dau$ collisions (closed circles). 
    This is compared to PHENIX data in $|y|<0.35$~\cite{ref:phenixdau_run8} (
closed diamonds). 
}
\label{fig:ptspectrum}
\end{center}
\end{figure}

The $\jpsi$ invariant cross section in $\pp$ collisions at $\snn = 200 \ \gev$
 and the $\jpsi$ invariant yield in $\dau$ at $\snn = 200 \ \gev$ are shown 
as functions of $\pt$ in 
Fig.~\ref{fig:ptspectrum}.
The $\pt$ spectrum in $\pp$ (left panel) extends the full STAR $\pt$ coverage 
to $0 < \pt < 14 \ \gevc$ ~\cite{ref:starauau} 
    and is consistent with previously published data from PHENIX~\cite{
ref:phenixpp_pol} at much smaller acceptance $|y|<0.35$. 
    The data are compared to the 
    color evaporation model (CEM)~\cite{ref:cem_star2} for prompt $\jpsi$ 
production in $\pp$ collisions. 
The CEM is able to describe the data well for the entire range of transverse 
momentum, 
    while it does not include contributions from $B$ decay which are expected 
to be $10-25\%$ for $\pt > 4 \ \gevc$ and decreasing at lower $\pt$~\cite{
ref:starauau}. 
		The model includes feed-down from heavier charmonium states ($\chic$ and $
\psiprim$),
		which are expected to contribute up to $40\%$ of the produced $\jpsi$ 
yield~\cite{ref:feeddown3}. 
The data are also compared to 
the non-relativistic quantum chromodynamics (NRQCD) calculations at next-to-
leading order (NLO) 
with color singlet and color octet (CS+CO) contributions~\cite{ref:nrqcd} for 
prompt $\jpsi$ production in $\pp$ collisions for $\pt > 4 \ \gevc$.
The model describes the data within large uncertainties, although it does not 
include contributions from $B$ meson decays. It also includes feed-down from $
\chic$ and $\psiprim$.
The color glass condensate (CGC) NLO CS+CO NRQCD model~\cite{ref:cgcnrqcd} 
for prompt $\jpsi$ for $\pt < 5 \ \gevc$ also describes the data within 
sizeable uncertainties.
    
The $\jpsi$ $\pt$ spectrum in $\dau$ measured by STAR (right panel) compared 
to PHENIX data taken at $|y|<0.35$~\cite{ref:phenixdau_run8} shows 
consistency within present statistical and systematic uncertainties. The 
resulting slope difference is consistent with zero within these uncertainties.

The integrated cross section for $\jpsi$ production in $\pp$ collisions for $|
y|<1$ at STAR has been calculated using the 
low-$\pt$ STAR data for $\pt < 2 \ \gevc$ combined with the previously 
published high-$\pt$ data for $\pt > 2 \ \gevc$~\cite{ref:starauau} and is 
found to be:
    
    \begin{equation}
    B_{ee}\frac{\mathrm{d}\sigma_{J/\psi}}{\mathrm{d}y} = 38 \pm 11 \ \textrm{
(stat.)} \pm 16 \ \textrm{(syst.)} \ \nb.
    \end{equation}    
    where the systematic uncertainty includes the uncertainty on the 
inelastic cross section in $\pp$ of $8\%$.
		
PHENIX data at low $\pt$ ($\pt < 2 \ \gevc$) have smaller statistical 
uncertainties compared to STAR measurements and therefore were used as a 
baseline for $\rda$. 
The integrated cross section was also re-calculated using the PHENIX data for 
$\pt < 2 \ \gevc$.
The value is shown in Eq.~\ref{eq_xsec} and is consistent with the STAR 
result within uncertainties.
		
		 \begin{equation}
    B_{ee}\frac{\mathrm{d}\sigma_{J/\psi}}{\mathrm{d}y} = 42.5 \pm 1.4 \ 
\textrm{(stat.)} \pm 4.8 \ \textrm{(syst.)} \pm 3.1 \ \textrm{(glob.)} \ \nb. 
    \label{eq_xsec}
    \end{equation}

The $\pt$ spectra provide valuable information about the $\jpsi$ production 
mechanism and $\jpsi$ interaction with the nuclear medium. The $\pt$ 
distribution is broadened in $\aaa$ and $\dau$ with respect to $\pp$ probably 
due to the Cronin effect, which arises from the multiple parton scattering in 
the initial state~\cite{ref:satz_pt2}. 
This broadening can be described by the formula: $\meanptt_{AA}=\meanptt_{pp}+
N^{AA}_c\delta_0$, where $N^{AA}_c$ is the average number of collisions for 
the projectile parton with target partons and $\delta_0$ is the average 
increase in $\pt$ a parton receives per collision. 
By comparing $\meanptt$ in different collision systems ($\pp$, $\pa$, $\aaa$) 
the parameter $\delta_0$ can be obtained.
Moreover, the analysis of this $\meanptt$ broadening in $\aaa$ may allow 
further study of the $\jpsi$ production mechanism. 
It is expected~\cite{ref:regen} that $\jpsi$ produced primarily by 
regeneration mechanism, will be characterized by a softer $\pt$ spectrum (
small $\meanptt$), while the direct $\jpsi$ from the initial hard scattering 
will show a rather hard $\pt$ spectrum (large $\meanptt$).
 Measurements of $\meanptt$ in $\pp$ and $\dau$ collisions serve as a 
baseline for such study allowing us to extract $\delta_0$. If the observed $
\meanptt$ is smaller in $\aaa$ collisions than expected from Cronin effect 
only, this may indicate that $\jpsi$ regeneration is contributing to the 
overall production.

\begin{figure}[hbt!!!!!!!!!!!]
\begin{center}
\includegraphics[width=.5\textwidth]{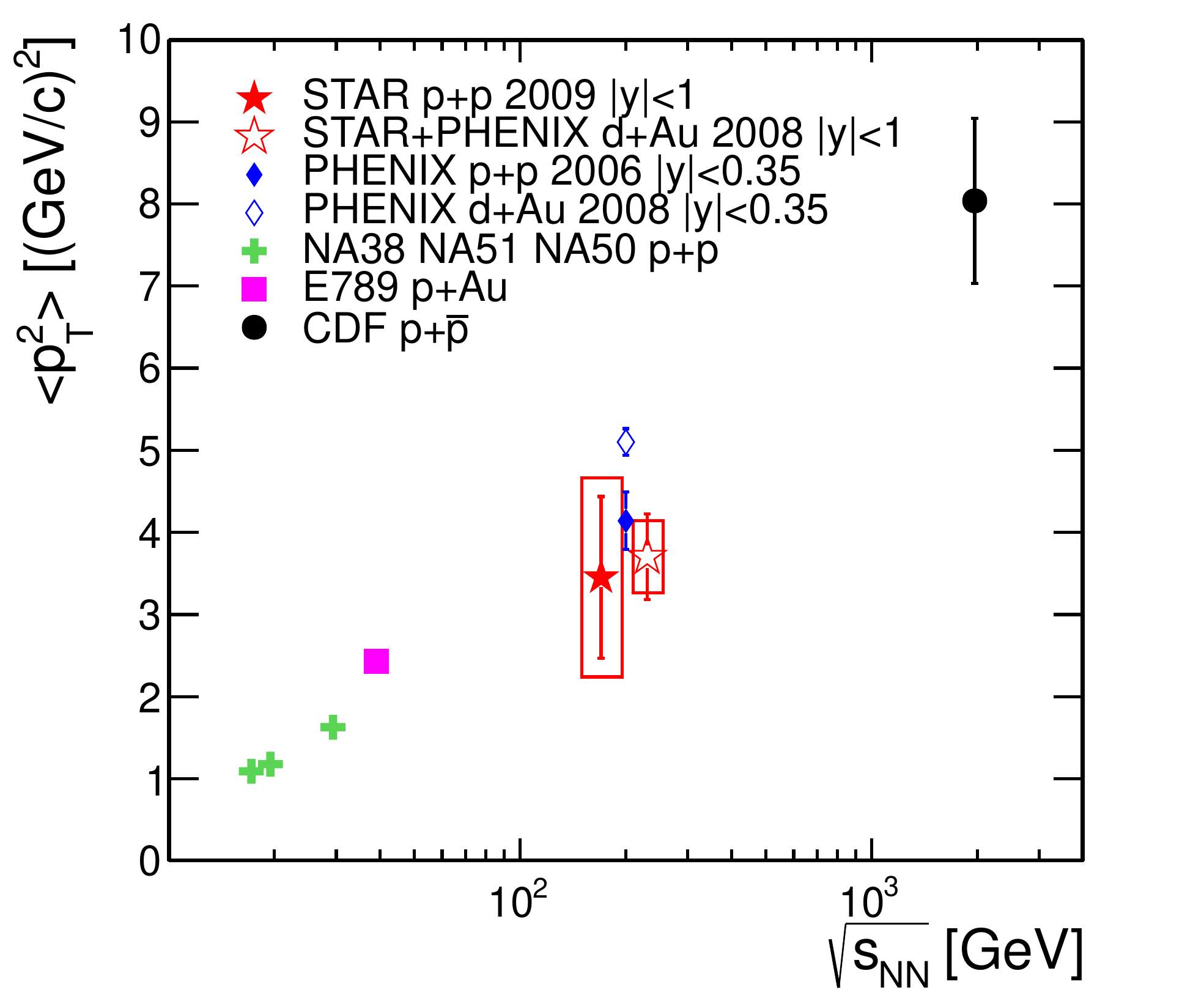}  
\caption{
  (Color online) The energy dependence of $\jpsi$ $\meanptt$. The STAR data (
closed stars for $\pp$ and open stars for $\dau$) are compared to PHENIX data~
\cite{ref:pt2_phenix,ref:phenixdau_run8} at $\snn = 200 \ \gev$ 
and various measurements at other collisions energies from the NA38, NA51, 
NA50, E789, and CDF experiments~\cite{ref:pt2_na50,ref:pt2_e789,ref:pt2_cdf}. 
The STAR results are shifted in $\snn$ for clarity.
}
\label{fig:ptsquare_pp}
\end{center}
\end{figure}

The $\jpsi$ invariant cross section from STAR has been used to study the $
\jpsi$ $\meanptt$ in $\pp$ and $\dau$ collisions at $\snn = 200 \ \gev$ shown 
in Fig.~\ref{fig:ptsquare_pp}.

The $\jpsi$ $\meanptt$ in $\pp$ collisions for $\pt < 14 \ \gevc$ was 
obtained directly from the STAR data, and its value is $\meanpttresult$. 
The results are consistent with PHENIX data in $|y|< 0.35$ at the same energy~
\cite{ref:pt2_phenix}. 
In $\dau$, the $\meanptt$ was calculated directly from combined STAR data 
points for $\pt < 3 \ \gevc$ and PHENIX data for $3 < \pt < 15 \ \gevc$. 
The $\meanptt$ was found to be $\meanpttresultdau$. 
The data are compared to various measurements at other collision energies 
obtained from the NA38, NA51, NA50~\cite{ref:pt2_na50}, 
E789~\cite{ref:pt2_e789} and CDF~\cite{ref:pt2_cdf} 
    experiments in Fig.~\ref{fig:ptsquare_pp}, and an increase of $\jpsi$ $
\meanptt$ with collision energy is observed. 
    Our measurements are consistent with the world data trend.

\begin{figure}[h]
\begin{center}
\includegraphics[width=.5\textwidth]{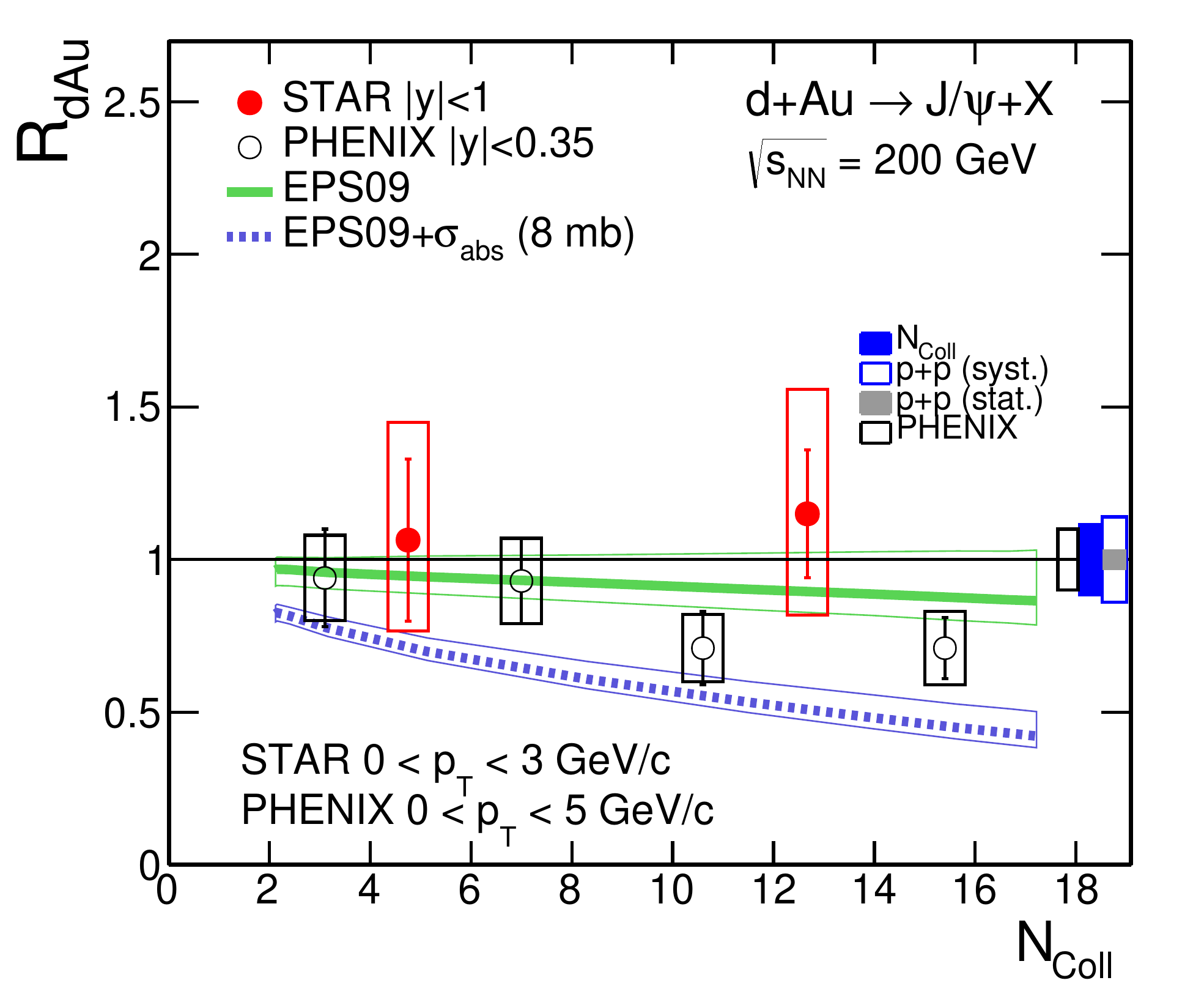} 
\caption{
  (Color online) The nuclear modification factor versus $\ncoll$ for $\jpsi$ 
with $|y|<1$ and $\pt < 3 \ \gevc$ in $\dau$ collisions (closed circles). 
  The central green line represents the predicted shadowing based on the EPS09
 nPDFs at next-to-leading order (NLO)
	~\cite{ref:RamonaEPS09, ref:EPS09} while the purple line shows shadowing 
combined with $\sigabs = 8.0\ \mb$, and the band indicates the uncertainty on 
the calculations. The data are compared to PHENIX results in $|y|<0.35$ ~\cite
{ref:phenixdau_run3} (open circles). 
}
\label{fig:raa_ncoll}
\end{center}
\end{figure}

The PHENIX $\pp$ data for $\pt < 2 \ \gevc$~\cite{ref:phenixpp_pol} and STAR 
data for $\pt > 2 \ \gevc$~\cite{ref:starauau} are combined as a $\pp$ 
baseline to provide better precision for $\rda$.
The $\pt$-integrated nuclear modification factor for $\jpsi$ with $\pt < 3 \ 
\gevc$ and $|y|<1$ is shown in Fig.~\ref{fig:raa_ncoll}. 
    The normalization uncertainty from the statistical and systematic 
uncertainty on the $\jpsi$ $\pp$ cross section, the uncertainty on the 
inelastic cross section, 
    and the uncertainty on $\ncoll$ are indicated on the vertical axis. 
The STAR results are consistent with unity within the uncertainties.

\begin{figure}[h]
\begin{center}
\includegraphics[width=.5\textwidth]{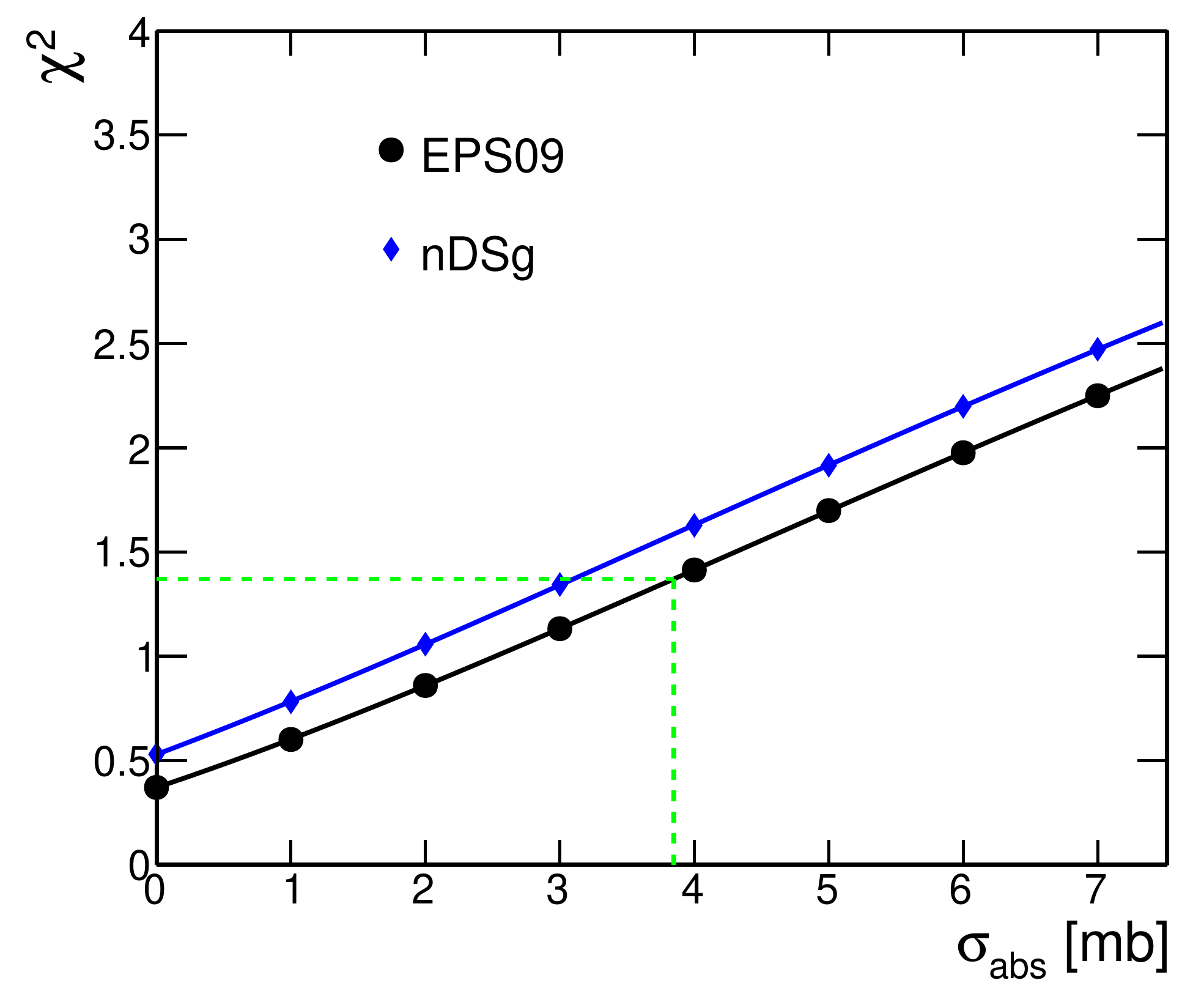}  
\caption{
  (Color online) The $\chi^{2}$ of the model calculations fitted to the STAR $
\jpsi$ $\rda$ using the EPS09~\cite{ref:EPS09, ref:RamonaEPS09} 
	and nDSG~\cite{ref:nDSg,  ref:RamonaNoEPS09} nPDFs as a function of a $\jpsi
$ absorption cross section $\sigabs$.
		The green dashed vertical and horizontal lines show the uncertainty on 
the $\sigabs$ and minimum $\chi^{2}$+1 respectively for the EPS09.
}
\label{fig:chi2}
\end{center}
\end{figure}

    The $\jpsi$ nuclear modification factor for $\dau$ has been compared to 
model calculations for cold nuclear matter effects on $\jpsi$ production in $
\dau$ collisions. 
    The CNM effects include the modification of nuclear parton distribution 
functions obtained from the 
    EPS09~\cite{ref:EPS09}
     and nDSg~\cite{ref:nDSg} parametrizations as well as effective $\jpsi$ 
absorption cross section ($\sigabs$)~\cite{ref:RamonaEPS09, ref:RamonaNoEPS09}
. 
    The absorption cross section was obtained by treating it as a free 
parameter in a $\chi^{2}$ minimization fit of the model calculations
    including CNM effects to the data. 
    The $\chi^{2}$ from the fit between the STAR data and model calculations 
as a function of the absorption cross section is shown in Fig.~\ref{fig:chi2} 
    for the EPS09
     and nDSg calculations of the nPDFs. The absorption cross section of $
\sigabsresult$ was obtained from the minimum $\chi^{2}$ value between the 
data and EPS09, which yields moderate $\chi^{2}$ compared to
     nDSg. By taking the minimum $\chi^{2}$+1 (green dashed line), a $3.8 \ 
\mb$ statistical and $2.1 \ \mb$ systematic uncertainty related to the 
fitting procedure was obtained.
	Due to the large uncertainties we quote an upper limit for nuclear 
absorption cross section of $\sigabs = 8.7 \ \mb$ at $2\sigma$ confidence 
interval. 
		The value for the absorption cross section is consistent with the results 
obtained using the 
		 nDSg parametrization and with other calculations performed at the same 
energy~\cite{ref:phenixdau_run3, ref:phenixdau_run8}. 
		The calculated $\rda$, assuming only shadowing and using the EPS09 nPDFs 
with the CTEQ6.1M free proton PDF at next-to-leading 
order (NLO)~\cite{ref:EPS09}, is shown in Fig.~\ref{fig:raa_ncoll} as the 
green solid line. The bands 
indicate the uncertainty from the EPS09 nPDFs. The model calculations agree 
with the data within the uncertainties. 
The EPS09 model calculations, with and without a nuclear absorption cross 
section of $\sigabs = 8.0\ \mb$ (within $1.84\sigma$ of our confidence 
interval) are also shown in Fig.~\ref{fig:raa_ncoll}.

\begin{figure}[h]
\begin{center}
\includegraphics[width=.5\textwidth]{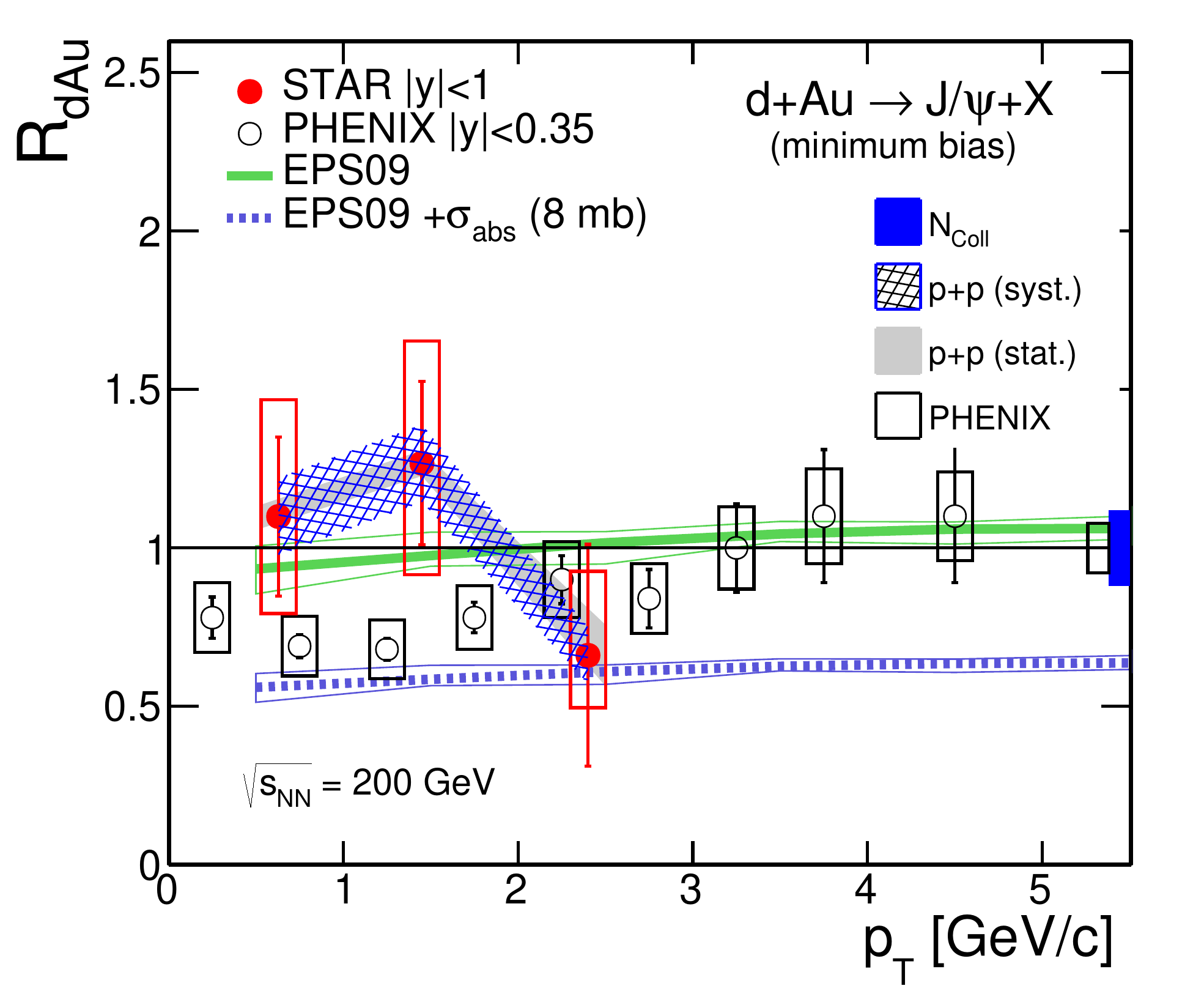}  
\caption{
  (Color online) The nuclear modification factor versus transverse momentum 
for $\jpsi$ with $|y|<1$ in 0-100\% central $\dau$ collisions (closed 
circles). 
  The central green line represents the calculated shadowing based on the EPS
09 nPDFs at next-to-leading order (NLO)~\cite{ref:EPS09} combined with a $
\jpsi$ absorption cross section of 
  $\sigabs = 0 \ \mb$~\cite{ref:RamonaEPS09} while the purple line shows 
shadowing combined with $\sigabs = 8.0\ \mb$, and the band indicates the 
uncertainty on the calculations. 
The data are compared to PHENIX data in $|y|<0.35$~\cite{ref:phenixdau_run8} (
open circles). 
}
\label{fig:raa_pt}
\end{center}
\end{figure}

The $\pt$ dependence of the $\jpsi$ nuclear modification factor in $|y|<1$ 
for $0-100\%$ centrality in $\snn = 200 \ \gev$ collisions 
is shown in Fig.~\ref{fig:raa_pt}. 
The gray band represents the statistical uncertainty on the measured $\jpsi$ cross section in $\pp$. The normalization uncertainties from the 
systematic uncertainty of 
the  $\jpsi$ $\pp$ cross section and the uncertainty of $\ncoll$ are 
indicated on the vertical axis. 
The results are compared to PHENIX data in $|y|<0.35$ and are in agreement 
within statistical and systematic uncertainties. 
The model calculations assuming shadowing only (EPS09) and shadowing combined 
with an absorption cross section $\sigabs = 8 \ \mb$ are also shown and both 
are consistent with the data. 
Note, that PHENIX results indicate suppression below $\pt$ of $2 \ \gevc$.


\section{Summary\label{sec:Summary}}

The production of $\jpsi$ within $|y|<1$ for $\pt < 4 \ \gevc$ in $\pp$ and $
\pt < 3 \ \gevc$ in $\dau$ collisions at $\snn = 200 \ \gev$, measured via 
the dielectron decay channel in the STAR detector, have been presented. 
The $\jpsi$ $\pt$ spectrum in $\pp$ collisions at STAR has been extended to 
cover $0 < \pt < 14 \ \gevc$ and has been used to calculate the $\jpsi$ $
\meanptt$ 
in $\pp$ collisions at $\snn = 200 \ \gev$. 
The results are consistent with other measurements at the same energy.
The obtained $\meanptt$ in $\dau$ of $\meanpttresultdau$ is consistent with 
the $\pp$ result of $\meanpttresult$ within large uncertainties, suggesting 
no significant Cronin effect.
The $\meanptt$ has also been compared to results 
from various experiments and exhibits an increase with increasing collision 
energy. 
The STAR data are consistent with the world data trend.
The modification of $\jpsi$ production in $\dau$ is consistent with no 
suppression within the measured uncertainties. 
The results have been compared to model calculations using the EPS09  
and nDSg parametrizations of the nPDFs including a J/psi nuclear absorption 
cross section as a free parameter.
An upper limit $\sigabs = 8.7 \ \mb$ within a $2\sigma$ confidence interval 
was obtained using the EPS09 calculations. 



\section{Acknowledgements}

We thank the RHIC Operations Group and RCF at BNL, the NERSC Center at LBNL, the KISTI Center in
Korea, and the Open Science Grid consortium for providing resources and support. This work was
supported in part by the Office of Nuclear Physics within the U.S. DOE Office of Science;
the U.S. NSF; the Ministry of Education and Science of the Russian Federation; NSFC, CAS,
MoST and MoE of China; the National Research Foundation of Korea; NCKU (Taiwan); 
GA and MSMT of the Czech Republic; FIAS of Germany; DAE, DST, and UGC of India; the National
Science Centre of Poland; National Research Foundation; the Ministry of Science, Education and
Sports of the Republic of Croatia; and RosAtom of Russia.


\bibliography{Bibliography}
\bibliographystyle{apsrev}

\end{document}